\documentclass[aps,amsmath,amssymb,twocolumn,longbibliography,floatfix]{revtex4-1}
\usepackage[utf8]{inputenc}
\usepackage{xcolor}
\usepackage{graphicx}
\usepackage{amsmath}
\usepackage{amsfonts}
\usepackage{amssymb}
\usepackage[caption=false]{subfig}

\begin{document}

\newcommand{\ds}{\text}

\title{Regularities and Irregularities in Order Flow Data}
\author{Martin Theissen}
\author{Sebastian M. Krause}
\author{Thomas Guhr}
\affiliation{Faculty of Physics, University of Duisburg-Essen, Duisburg, Germany}

\begin{abstract}
We identify and analyze statistical regularities and irregularities in the recent order flow of different NASDAQ stocks, focusing on the positions where orders are placed in the 
orderbook. This includes limit orders being placed outside of the spread, inside the spread and (effective) market orders.
We find that limit order placement inside the spread is strongly determined by the dynamics of the spread size.
Most orders, however, arrive outside of the spread.
While for some stocks order placement on or next to the quotes is dominating, deeper price levels are more important for other stocks.
As market orders are usually adjusted to the quote volume, the impact of market orders depends on the orderbook structure, 
which we find to be quite diverse among the analyzed stocks as a result of the way limit order placement takes place.
\end{abstract}

\maketitle

\section{Introduction}

The progress in information technology always has a huge impact on stock markets and related systems. It affects not only the trading process itself, but also the availability of data, the tools used for their analysis and the model building that follows. The accessibility of detailed order flow data capturing order book dynamics is useful for researchers and practitioners. 
Apart from an understanding of financial markets, long term research questions concern systemic issues, such as the usefulness of (de)regulations of financial markets \cite{farmer2009economy}, and the consequences of high
frequency trading for market stability \cite{lee2013microstructure, kirilenko2015flash, brogaard2010high}. Practitioners investigate the profitability of trading strategies \cite{brogaard2010high}. From both viewpoints, systemic and practical, an improvement of agent based models \cite{patzelt2013inherent, meudt2016equilibrium, krause2013spin} exploiting order flow analysis is desirable. 

While stylized facts of price time series are entrenched for decades \cite{cont2001empirical}, comparable knowledge on the level of order book dynamics has not yet been established, for a comprehensive review see \cite{GouldLOBreview}. As the variability of results is at least partly due to the non-stationarity of markets, empirical findings have to be steadily reviewed with respect to their validity for contemporary data \cite{chakrabarti2016absence}. 
For example, in NASDAQ stocks in 2002 large portions of all limit orders are placed far from the quotes \cite{Potters02}. In contrast, later studies find a dominating role of the quotes and suggest queue models for characterizing the execution sequence of limit orders \cite{gareche2013fokker}. If there are substantial differences in the order flow of different stocks at the same time in the same market, this has implications for the description of the market on the systemic level. When analyzing stock interactions \cite{Wang2016, hasbrouck2001common, boulatov2013informed, pasquariello2013strategic,chordia2000commonality}, one has to be aware of heterogeneous order flow mechanisms for different stocks. This aspect was not sufficiently investigated so far, maybe partly because the movement of the market has been found to be dominated by collective effects \cite{stepanov2015stability}. 

Here we analyze order flow data of $96$ NASDAQ stocks in early 2016. We focus on where orders are placed in the order book, and compare statistical regularities across stocks. We find a rich variety of spread widths, order placement measures and price returns caused by single market orders. Further we search for a clustering of stocks, grouping together stocks with similar behavior. 

This study is organized as follows: In Sec.~\ref{sec:data} we describe the data. In Sec.~\ref{sec:results} we reconstruct market orders from the data, and investigate the prices at which limit orders enter the order book. These prices help us to cluster the stocks into groups with similar behavior. The significance of this grouping with respect to liquidity measures as the spread size, number of orders on the quotes, and share of market orders among all orders is investigated. We find striking differences for the returns caused by single market orders for different groups of stocks. Finally, in Sec.~\ref{sec:summary} we summarize our results and give an outlook.

\section{Data}\label{sec:data}

We analyze the data Historical TotalView-ITCH from NASDAQ US. 
We group the data into the order flows of each stock on one specific day. For example \ds{20160307\_AAPL} comprises of the orders for Apple shares (the ticker is AAPL) on the NASDAQ US on March, 7 2016 (a Monday). 
Our data comprise five days from March, 7 2016 to March, 11 2016. Out of the 100 stocks listed in the NASDAQ 100 in this period \cite{wikiNasdaq100}, four stocks are not available in \cite{tp}, and therefore cannot be included in our analysis. 
The data contain information about limit orders being placed,
deleted, partially canceled, partially traded and fully traded.
Moreover, they contain 
information about trades against hidden orders. A detailed description of the data can be found in \cite{huang2011lobster}.

We analyze data from times between 10:00 am and 3:30 pm (New York time). 
These are the regular trading times of the NASDAQ excluding the first and last $30$ minutes.
We neglect these, because order flow dynamics at the opening and closing of the market have different statistical properties. All events have a time stamp in milliseconds.
Events happening in the same
millisecond have the same time stamp, although they may not have happened simultaneously. 
Chronological order is maintained, as incoming orders are processed by the market in the same succession as in which they arrive.

Our data reveal several characteristics of limit orders,
order book dynamics and trades. However, the data only shows the net effect of
a trader's action in terms of simple limit orders and trades. 
Exotic orders appear in a rich variety, making it impossible to fully reconstruct them from the order flow \cite{order_types}. 
The only exception are market orders.
These will be reconstructed from successive trade events due to their high relevance for 
our research goal.  
This can of course only be done approximately. 
In our reconstruction, we can not distinguish real market orders from effective market orders, \textit{i.e.} limit orders crossing the spread. 
The data was tested to be free of internal contradictions, as also reported in other works \cite{howtohft_blog,caltech}.

\section{Results}\label{sec:results}

In Sec.~\ref{sec:reconstruction} we perform the reconstruction of market orders, as one market order can trigger many consecutive trades, and this information is not provided in the data. Further we neglect market orders which are solely traded against hidden limit orders. The share of market orders among all orders is calculated separately for different stocks, as well as similar measures characterizing limit orders. In Sec.~\ref{sec:in_spread} we discuss statistical regularities of in-spread limit orders and use them to perform a clustering of stocks into groups with similar behavior. We test in how far the stocks within the same group also show similar properties according to order counts per day. In Sec.~\ref{sec:off_spread} we see that the grouping of stocks according to in-spread limit orders is also relevant for off-spread limit orders. Finally, in Sec.~\ref{sec:market_orders}, we analyze the impact of single market orders, again considering the grouping found for in-spread limit order placements.

\subsection{Market Order Reconstruction and General Statistics}\label{sec:reconstruction}

Our data do not explicitly include market orders, thus we have to reconstruct them from the trade events. Our reconstruction modifies the approach of \cite{Hautsch2011-2}.
We trace subsequent trades to a single market order if all limit orders possibly traded against this market order are of the same type (for example sell). For trades against hidden limit orders, the type is unknown, so we include such trades in the market order reconstruction without discriminating sell or buy type. Apart from that we require that events of a different type do not come in between the trades, as for example a limit order arrival or cancellation, and that partially traded limit orders can only be the last event. 
Finally we require that all trade events take place within a time window of fixed length. According to the market specification \cite{Nasdaq_execution_time}, the order ``execution time is less than one millisecond.''
This statement suggests that a consolidation of trade events
should be only taken into account as long as the time stamp difference between
the first and the last trade of a market order is at most one millisecond.
We call the number of trades triggered by one market order ``cluster size''. This terminology (``cluster'' denoting a group of trades) discussed here has nothing to do with the clustering discussed in the upcoming chapters.

\begin{figure}[htb]
		\includegraphics[width=0.45\textwidth]{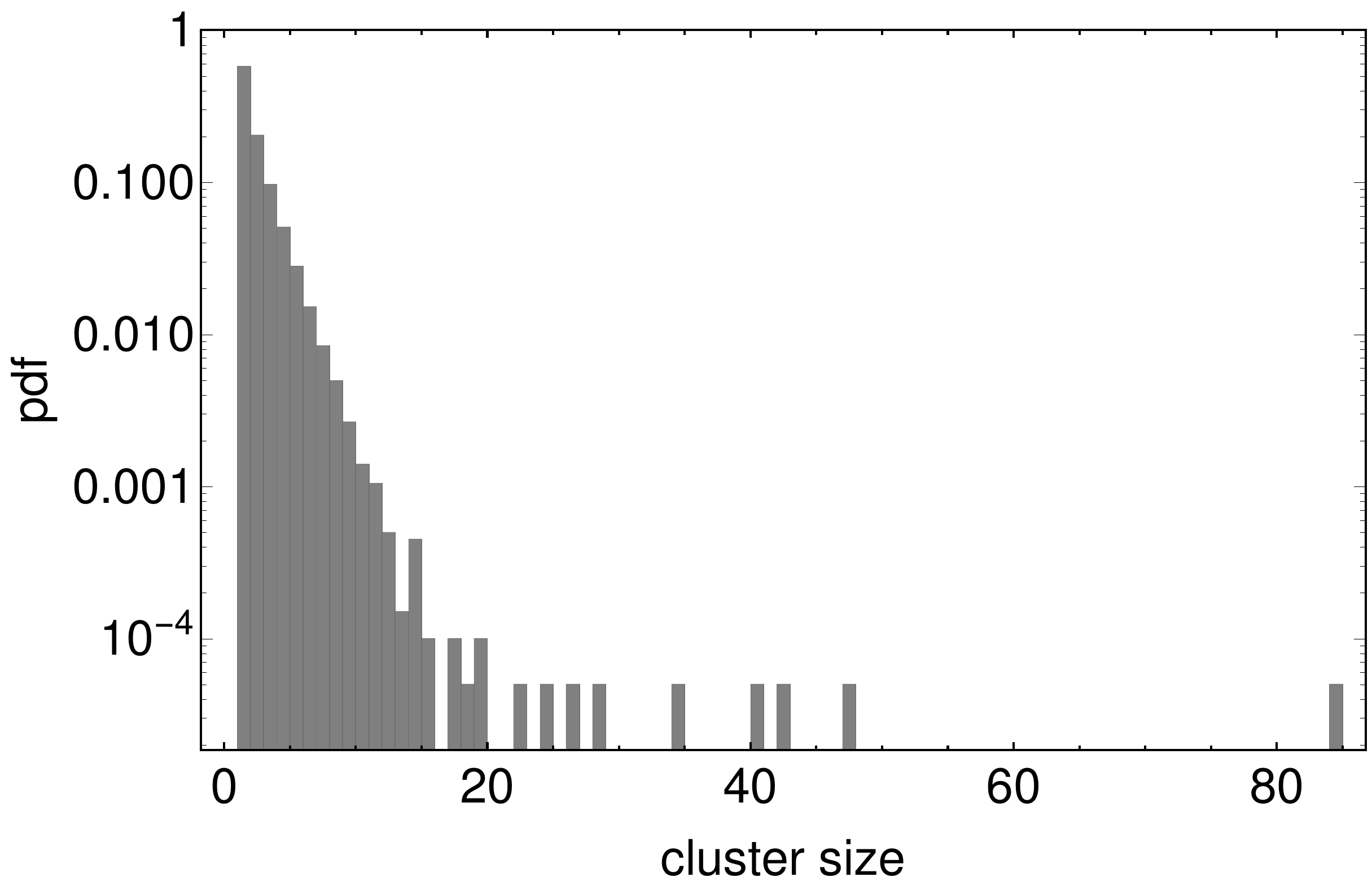}
    \caption{Distribution of the cluster size (number of trades consolidated to one market order) for \ds{20160307\_AAPL}}
	\label{fig:cluster_size_dist}
\end{figure}

Figure \ref{fig:cluster_size_dist} shows the distribution of these cluster sizes for \ds{20160307\_AAPL}.
We find that $58\%$ of all market orders result in only one single trade. However, there are also groups of up to more than 80 trades against one market order. We checked the robustness of our reconstruction with respect to the maximum time difference allowed between the first and the last trade of a market order. There are only minor changes in the number of market orders, even if we increase the maximum time window from one millisecond to half a second. The respective changes in the distribution of cluster sizes are barely visible compared to the distribution in Fig. \ref{fig:cluster_size_dist}. For that reason the distributions for other maximum time windows are not shown here.

Moreover, for every group of $N$ trade events possibly considered due to one market 
order, between one and $N$ market orders might have triggered this series of events. This problem can not be unambiguously solved with help of our data. 
However, as we will see in the sequel, given that the number of potential market orders is much smaller than the number of limit order 
insertions and deletions, we may assume that two market orders arriving subsequently without being interrupted by a limit order placement or deletion is a very rare incident. Thus, if allowed by the other constraints, we 
cluster all multiple trade events to one market order.

There are market orders that can not be assigned 
to either buy or sell type. This happens for trades executed exclusively against hidden orders.
In the upcoming study, we have to dismiss these market orders.
We calculate the frequency of undirected market orders among all market orders for the order flow of one stock on one day, and repeat this for all 96 stocks on all five days, resulting in a number of 480 relative frequencies. We find that the proportion of undirected market orders among all market orders is on average around $12\%$. 
\begin{figure}[htbp]
		\includegraphics[width=0.45\textwidth]{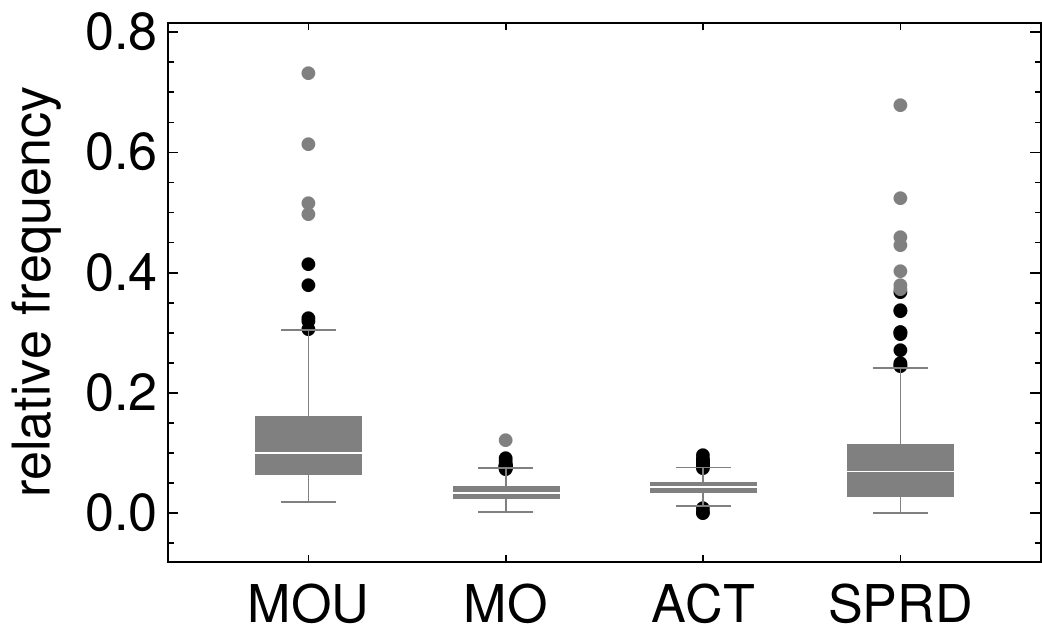}
\caption{Box-Whisker-Plots for the relative frequencies of undirected market orders among all market orders (MOU), market orders among all orders (MO), active limit orders among all limit orders (ACT) and in-spread limit orders among all limit orders (SPRD).}
\label{fig:summary_statistic}
\end{figure}
The scattering of this quantity over the different data sets is presented with a Box-Whisker plot  in Fig. \ref{fig:summary_statistic} on the left (MOU). The white horizontal line in the gray box indicates the above mentioned average of $12\%$. The gray box indicates both centered quartiles, so we see that half of the data is concentrated within the range $6.6\%$ to $16.1\%$. The antennas reach up to the largest (smallest) data point within 1.5 times the interquartile range added (subtracted) to the upper (lower) quartile. The additional dots outside of the antennas correspond to the outliers: There are a few data sets in which  a certain stock had a particularly high frequency of undirected market orders on a certain day.
Altogether we find that for most of the data sets, meaningful information can be extracted from the market orders. The rejected undirected market orders only constitute a small fraction of all market orders.

In the second Box-Whisker plot labeled with MO, we see that the relative frequency of market orders among
all orders is rather small
with an average of about $3.7\%$. Here, the regularity along our data sets is much stronger than it was for the relative amount of undirected market orders (MOU).
We checked that the volume stored in the order book does neither systematically increase nor decrease 
throughout the day.
Thus, either order volumes or limit order deletions have to compensate that the number of orders
providing liquidity, \textit{i.e.} limit orders, is much higher than the number of orders taking liquidity, \textit{i.e.} market orders.
We find that most limit 
orders are deleted. On average, only $4.4\%$ of all limit orders per data set are (at least partially) traded, cf. the Box-Whisker plot labeled as ACT. This means that the majority of limit orders is deleted without participating in the trading process. 
Trading is obviously {\it not} the predominant factor that clears orders from the order book. This is also known from earlier studies \cite{GouldLOBreview}. The relative frequency of orders that end up being deleted among all incoming orders is one of the quantities that clearly reflect non-stationarity of markets, see for example \cite{Bouchaud02}, where only $10\%$ of all limit orders are reported to be deleted (without at least partial execution) on the Paris stock exchange in 2002.

Finally, we are interested
in the relative frequency of in-spread limit orders among all limit orders. 
The mean relative frequency of in-spread
limit orders is $8.5\%$. The Box-Whiskers plot labeled SPRD shows that this relative frequency varies stronger within the data sets as compared to those frequencies discussed before. 
The most obvious reason to expect differences for this quantity among different 
data sets is the fact that the spread does not always allow in-spread limit orders.
We will return to that point in Sec.~\ref{sec:in_spread}. 

\subsection{In-spread Limit Orders}\label{sec:in_spread}

\begin{figure*}[htbp]
\centering
\subfloat[\ds{AAPL}]{
		\includegraphics[width=0.32\textwidth]{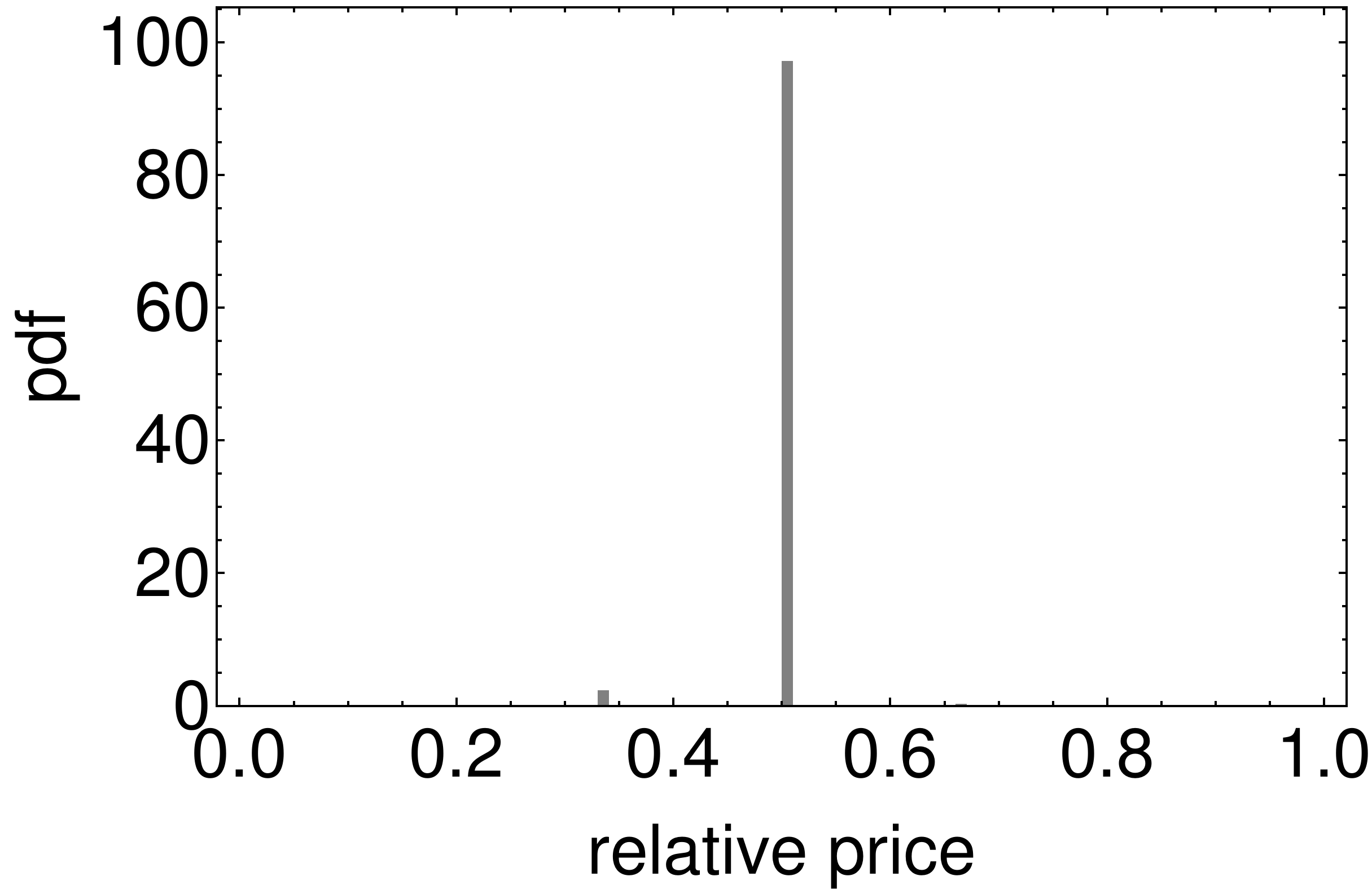}
		\label{fig:rp_sprd_hist_AAPL}}
\subfloat[\ds{CERN}]{
		\includegraphics[width=0.32\textwidth]{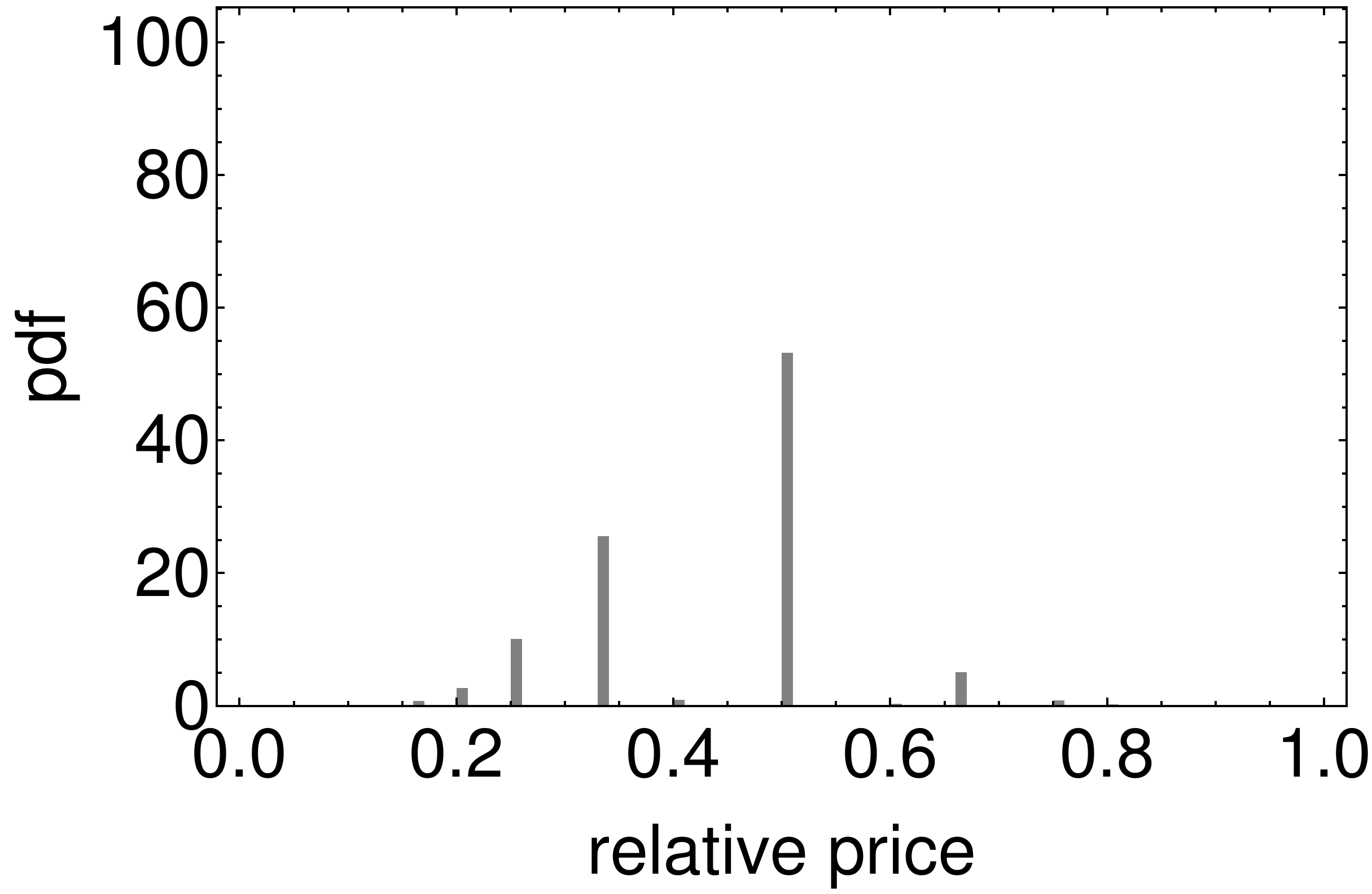}}
\subfloat[\ds{GOOG}]{
		\includegraphics[width=0.32\textwidth]{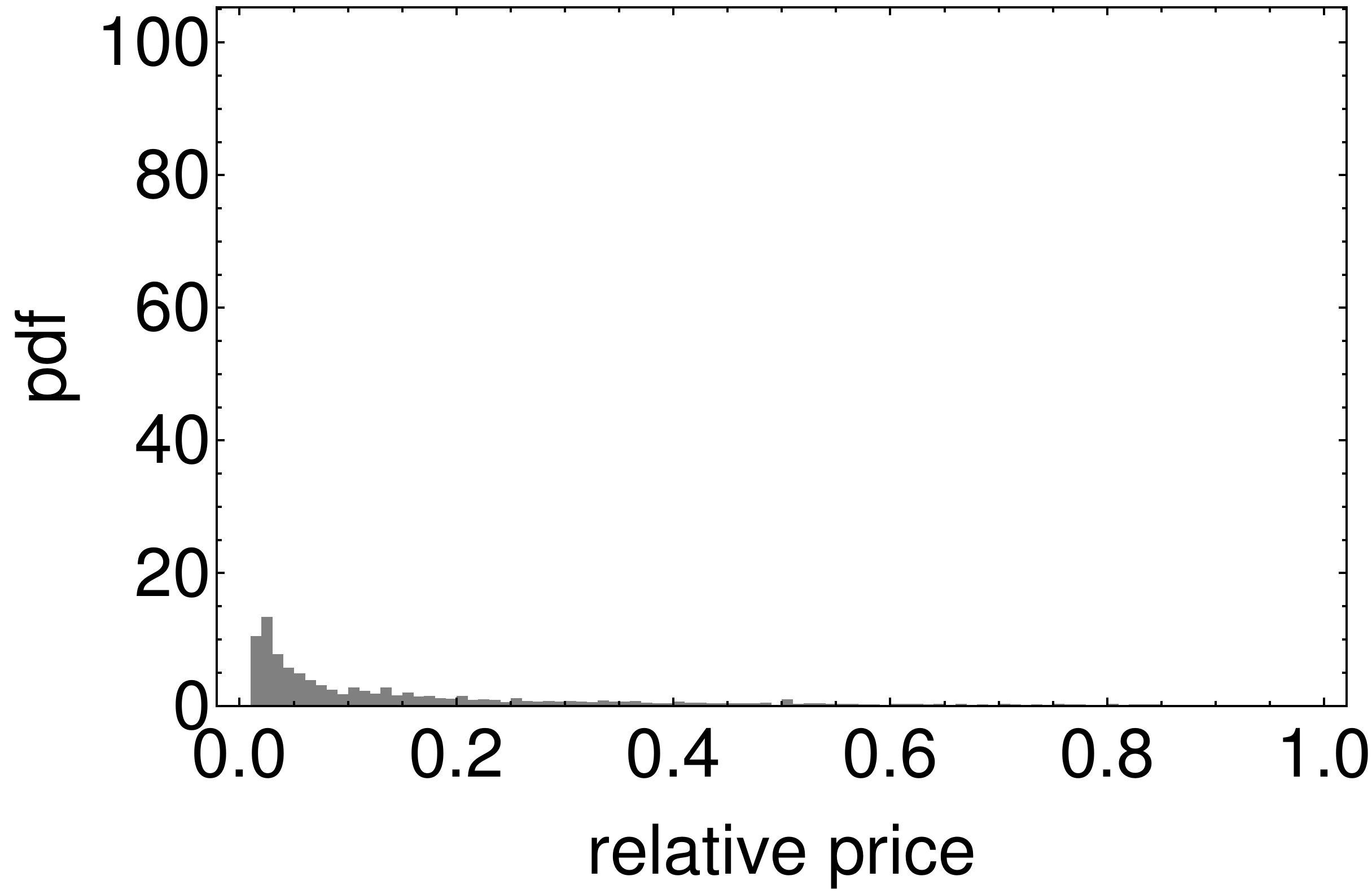}
		\label{fig:rp_sprd_hist_CERN}}
		\\
\subfloat[\ds{AAPL}]{
		\includegraphics[width=0.32\textwidth]{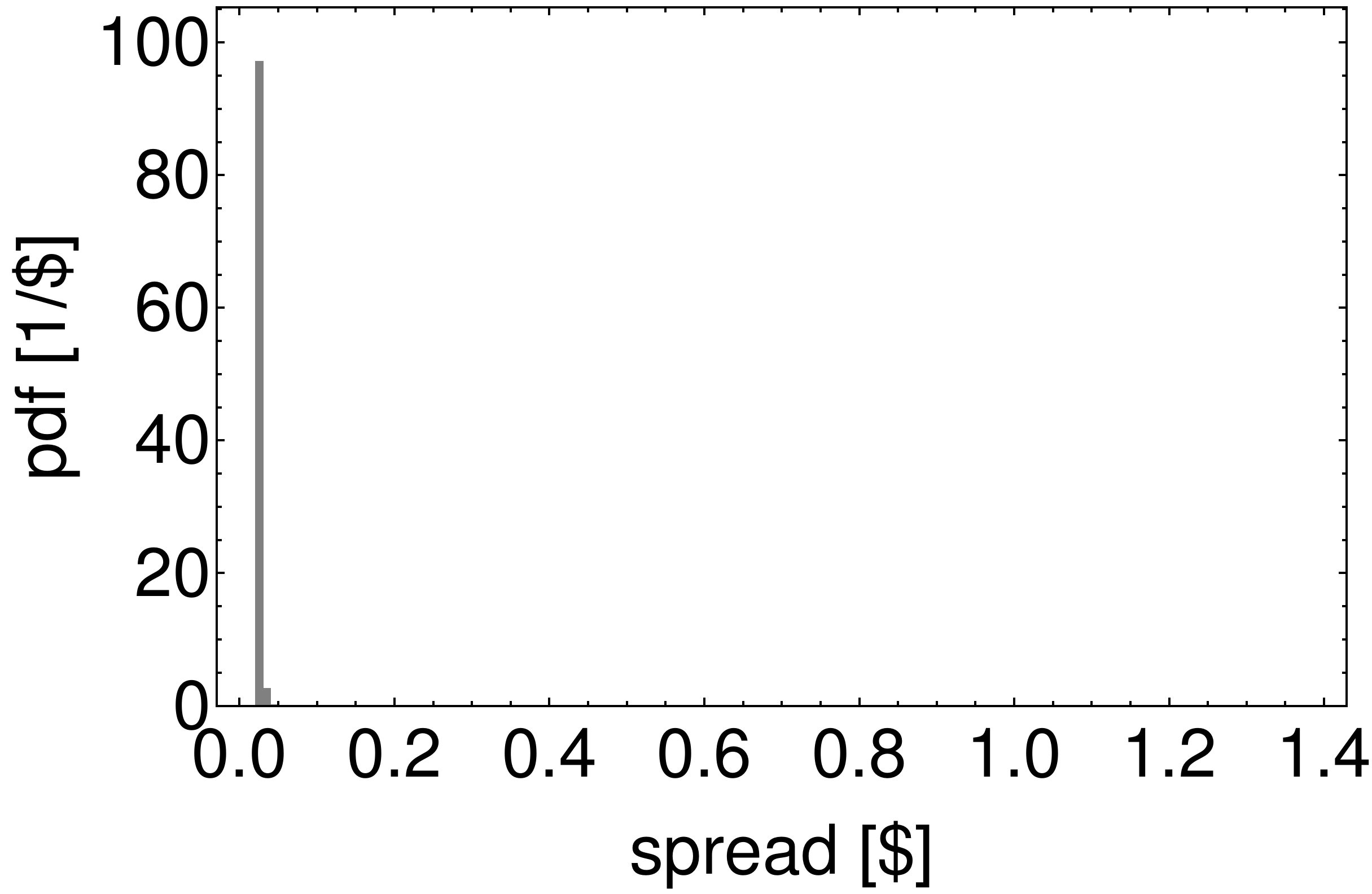}}
\subfloat[\ds{CERN}]{
		\includegraphics[width=0.32\textwidth]{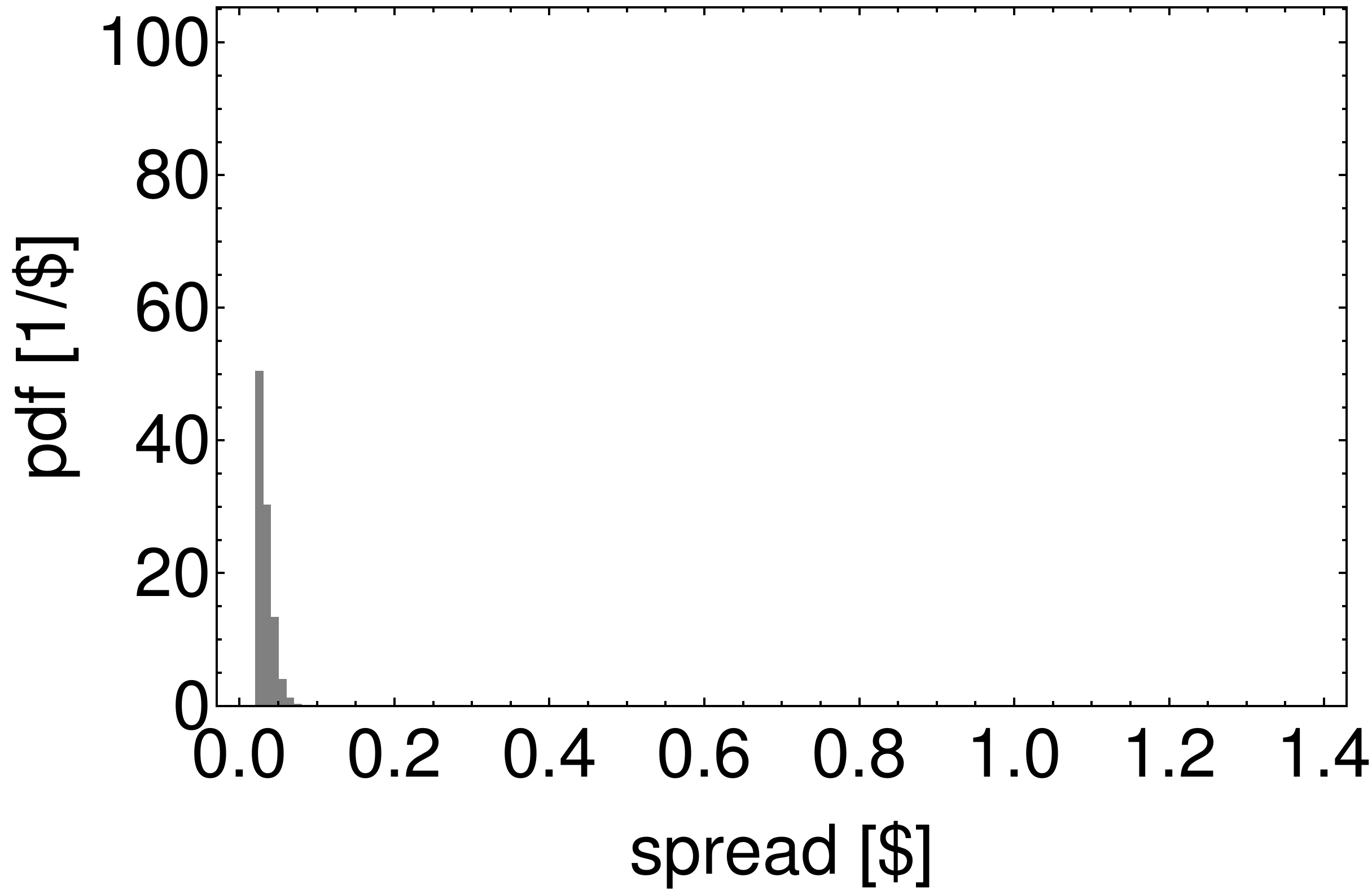}
		\label{fig:rp_sprd_hist_GOOG}}
\subfloat[\ds{GOOG}]{
		\includegraphics[width=0.32\textwidth]{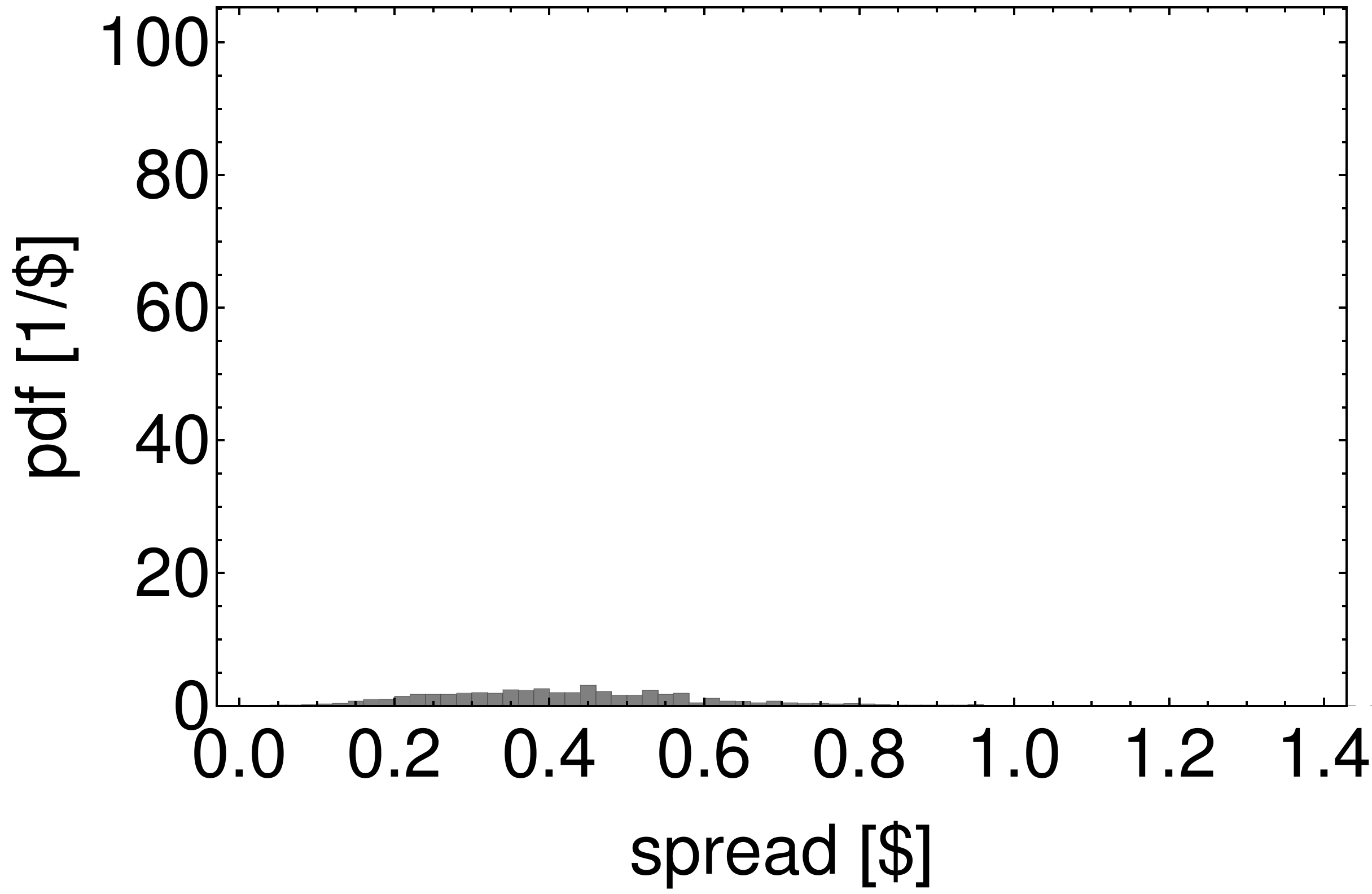}}
\caption{Distributions of relative prices of in-spread limit orders (abc) and spread sizes prior to in-spread limit orders (def) for three data sets in different clusters. We choose to work with the data sets
\ds{AAPL} ((ad), cluster C1), \ds{CERN} ((be), cluster C2) and \ds{GOOG} ((cf), cluster C4) on the day \ds{20160307}.}
\label{fig:rp_sprd_hist}
\end{figure*}
We analyze the aggressiveness of in-spread limit orders, \textit{i.e.} how close they are placed to the opposing quote. 
Hence, we evaluate the distribution of the relative price
\begin{align*}
\tilde p=\frac{p-p_q}{s},
\end{align*}
where $p_q$ denotes the price of the quote on the same side and $s$ denotes the
size of the spread, both at the time where the limit order is placed. 
For sell and buy limit orders, small $\tilde{p}$ implies a small aggressiveness of the order, moving the quote only slightly. 
We calculate for each of the 96 stocks the in-spread relative price distributions over all five days. Examples of such distributions are depicted in Fig. \ref{fig:rp_sprd_hist}(ace) for three stocks. The distribution for Apple (AAPL) has a dominant sharp peak at $\tilde{p}=1/2$. As we will see below, this is because the spread mostly only opens for one tick, and this one tick is then the only possible choice for placing an in-spread limit order. The relative price distributions are very diverse for different stocks. 

To find groups of stocks with similar behavior, we compare relative price distributions for pairs of stocks. A robust way to deal with mixed distributions containing sharp peaks is to use the cumulative distribution. Calculating the maximum over all possible values of the difference between two cumulative distributions, we obtain the Kolmogorov-Smirnov (KS) distance 
measure. As a result of the stationarity of these distributions over the five days of our observations, we can use the average distribution for each stock without loss of information. 
\begin{figure}[htbp]
\centering
		\includegraphics[width=0.47\textwidth]{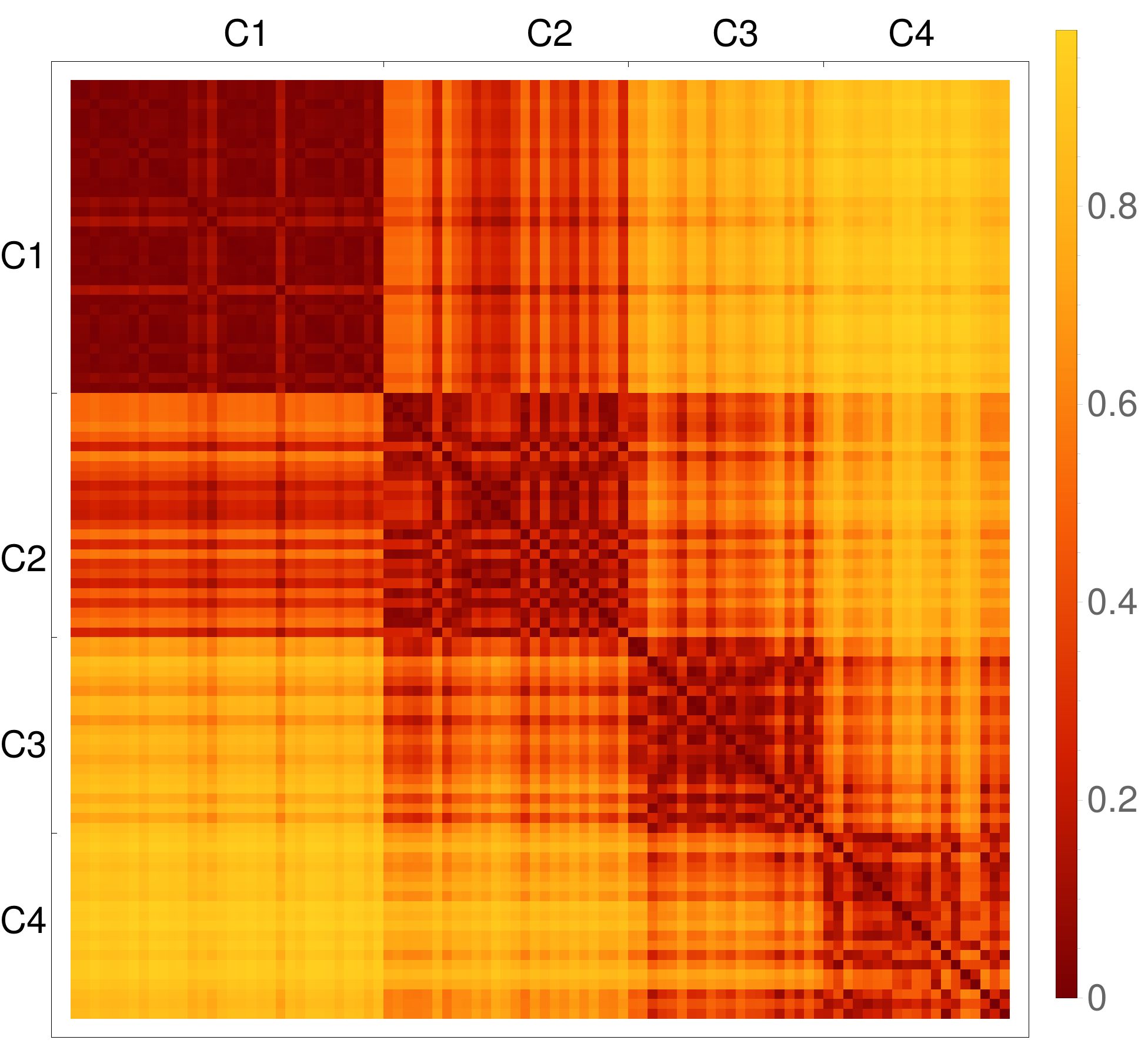}
\caption{Kolmogorov-Smirnov distance matrix of in-spread order relative price distributions of our 96 stocks. The stocks were rearranged so that the clustering structure is visible, with clusters denoted C1, C2, C3 and C4.}
	\label{fig:rp_sprd_merged_distmat}
\end{figure}
The KS distance is shown in  Fig. \ref{fig:rp_sprd_merged_distmat} for all pairs of stocks. The ordering of stocks is carried out with a clustering of stocks using the relational $k$-means algorithm \cite{Szalkai2013}, with the KS distances as an input. The relational $k$-means algorithm partitions the stocks into $k$ clusters, such that distances between stocks in the same cluster are small \cite{Szalkai2013}, and therefore stocks in the same cluster are similar. As a consequence, a pair of stocks with a large KS distance is most likely separated into different clusters. 
With help of the mean silhouette information criterion \cite{rousseeuw1987silhouettes}, we find that a number of four clusters is most appropriate to work with.
The sizes of the clusters are as follows: $33\%$ of the stocks belong to cluster C1, $21\%$, $26\%$ and $20\%$ of the data sets belong to the clusters C2, C3 and C4, respectively.

Stocks in the same cluster have similar distributions of relative prices. 
In fig. \ref{fig:rp_sprd_hist} we show distributions for representative stocks out of three of the clusters. 
As a result of the condition of times immediately before a limit order is placed in-spread, the minimum spread size observed is two ticks, although the spread size may equal only one tick for most of the time per day. 
As discussed above, the distribution shown in Fig.~\ref{fig:rp_sprd_hist}(a) is essentially a sharp peak at $0.5$. In Fig.~\ref{fig:rp_sprd_hist}(d) we see that the spread is mostly only two ticks wide, before a new limit order arrives inside the spread. 
We adjusted the bin size to the tick size of one cent. By multiplying the density with the bin width we find that the probability for having a spread of two ticks is 97\%. In this case, the relative price can only be $\tilde{p}=0.5$. Stocks in our first cluster typically have a small spread. Such stocks have also been described as large tick stocks \cite{eisler2012price,gareche2013fokker,dayri2015large}, because the ticks are large compared to the spread size which liquidity providers are willing to offer. Instead, there is a large queue of limit orders waiting on the quotes for being executed, and the order of execution (according to arrival times) is important for investment strategies \cite{gareche2013fokker}. To which extent this reflects in the order arrival outside of the spread in the different clusters found here will be analyzed in Sec.~\ref{sec:off_spread}.

Cluster C4 also has a clear interpretation, see the distribution of relative prices in Fig.~\ref{fig:rp_sprd_hist}(c). It has a maximum at the relative
price which corresponds to the second least aggressive price. The density is decreasing towards higher 
relative prices. The bin size is chosen bigger than the discretization. The corresponding distribution of spread sizes Fig.~\ref{fig:rp_sprd_hist}(f) shows that the spread is usually so large that relative prices are almost not constrained by the spread size.
Order placements with small aggressiveness are favored. A reason could be that traders want to have execution priority for their limit orders, but they want to stay close to the current quotes, because undercutting favors the liquidity taker. 
The behavior of stocks in the fourth cluster is closer to the market situation of older market studies \cite{Farmer2004,Mike2005}.
Interestingly, both types of market dynamics exist at the same time in the same market in 2016. The small tick stocks only comprise about $1/5$ of all stocks for the data we analyze.

Clusters C2 and C3 can be understood as originating from intermediate and varying spread sizes. The spreads of stocks in the intermediate clusters are located between the small spreads of stocks in cluster C1 with strong discretization effects, and the quasi continuous spreads of stocks in cluster C4. 
We consider cluster C2 as an example.
The data set for \ds{CERN} considered here is part of cluster C2, see Fig.~\ref{fig:rp_sprd_hist}(b). If the spread is only two ticks, the relative prices are $0.5$.
However, sometimes the spread is larger, allowing for other relative prices. The fact that we mostly observe relative prices smaller or equal to $0.5$ shows that generally less aggressive limit order placement is favoured over a more aggressive one.

In the literature, the distribution of relative prices in the spread is fitted by some analytical probability density \cite{Mike2005}. 
Although distribution fitting might work well for the cluster C4, the distributions for 
the other three clusters might not be well described that way. 
Eventually, not only trader behavior but also conditions set by the spread size determine the distributions of relative prices,
which makes a unifying characterization difficult.

\begin{figure}[htbp]
	\centering
		\includegraphics[width=0.45\textwidth]{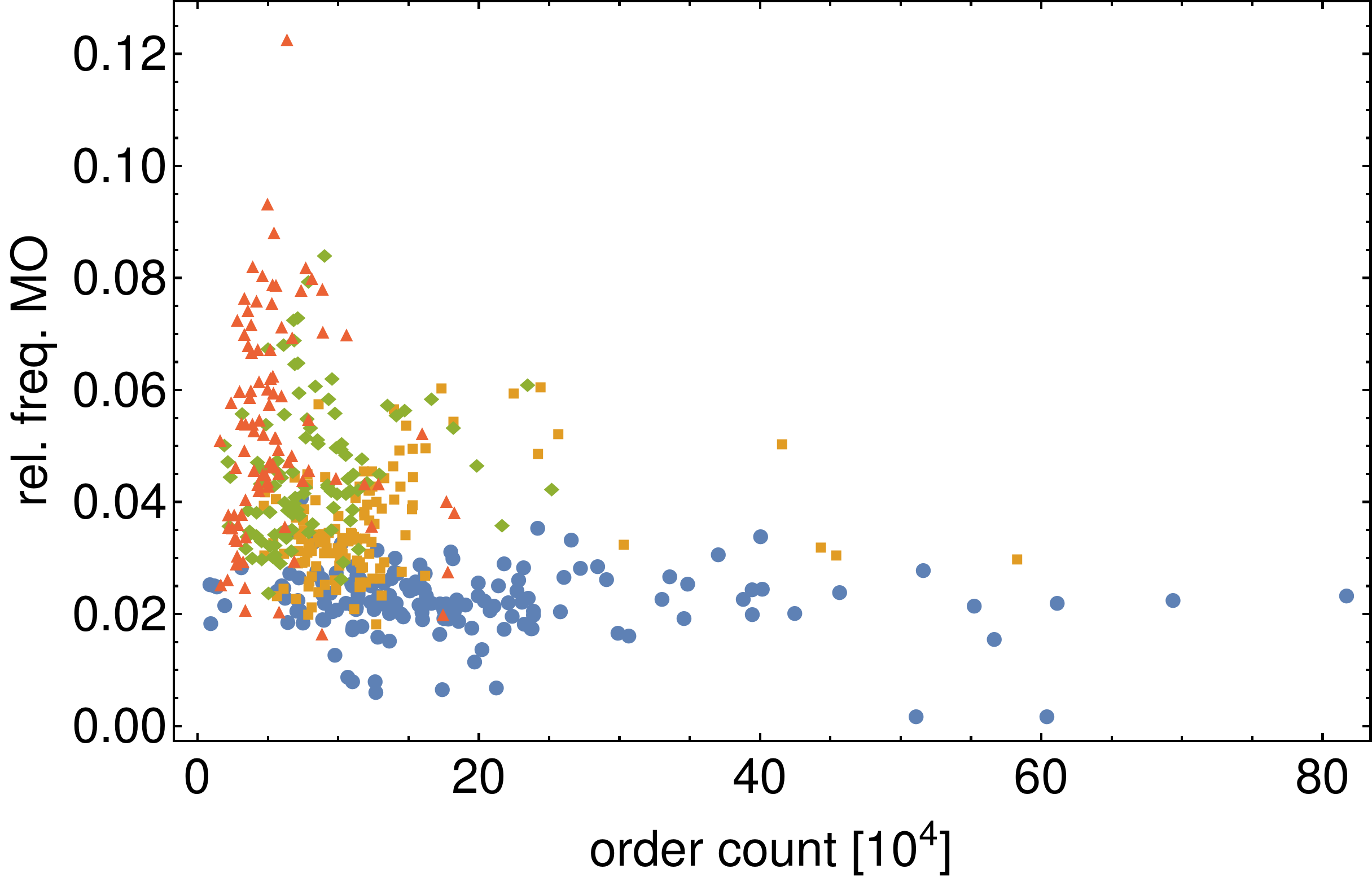}
\caption{Scatter plot of daily order counts for each data set vs. the relative frequency of market order among all orders. The colors and symbols indicate the affiliation of a stock to one of the four clusters C1 (blue circles), C2 (yellow squares), C3 (green diamonds) or C4 (red triangles).}
	\label{fig:rp_cluster_activity}
\end{figure}

Before we turn to the off-spread limit orders, we investigate to which extent the different characteristics relate to the overall market activity for a stock. Figure \ref{fig:rp_cluster_activity} shows the relative frequency of market orders among all orders plotted against the daily order counts for each data set in a scatter plot. The colors and symbols correspond to the different clusters.
The following regularity emerges: In cluster C1 (blue circles), where the typical spread is small, total order counts tend to be high, while 
the relative frequency of market orders among all orders is rather small. This hints at a strong presence of algorithmic trading 
in these data sets, since the algorithms typically imply a high throughput of limit orders \cite{harris2003trading,hendershott2011does}.
In contrast, stocks in cluster C4 (red triangles) with large spreads come hand in hand with small order counts and comparably high relative frequencies of market orders. This might be due to a weaker presence of algorithmic traders.

\subsection{Off-spread Limit Orders}\label{sec:off_spread}

\begin{figure}[htbp]
	\centering
		\includegraphics[width=0.45\textwidth]{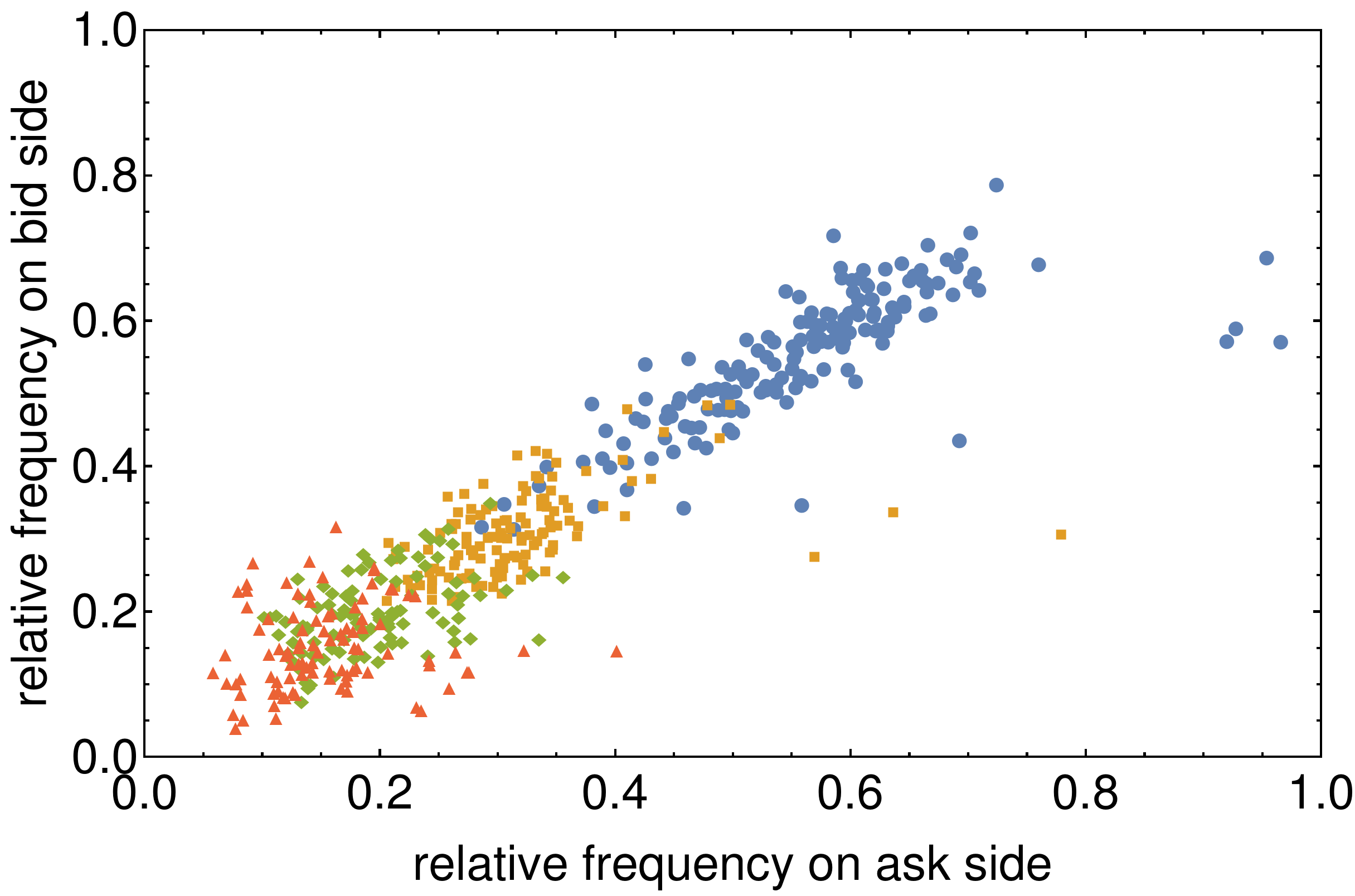}
\caption{Scatter plot of relative frequencies buy and sell of limit orders placed on the quotes. The 
colors represent the cluster to which the relative price distribution of the respective data set was assigned. Symbols and color coding as in Fig.~\ref{fig:rp_cluster_activity}.}
	\label{fig:rp_cluster_placement0}
\end{figure}
\begin{figure*}[htbp]
\centering
\subfloat[\ds{AAPL}]{
		\includegraphics[width=0.45\textwidth]{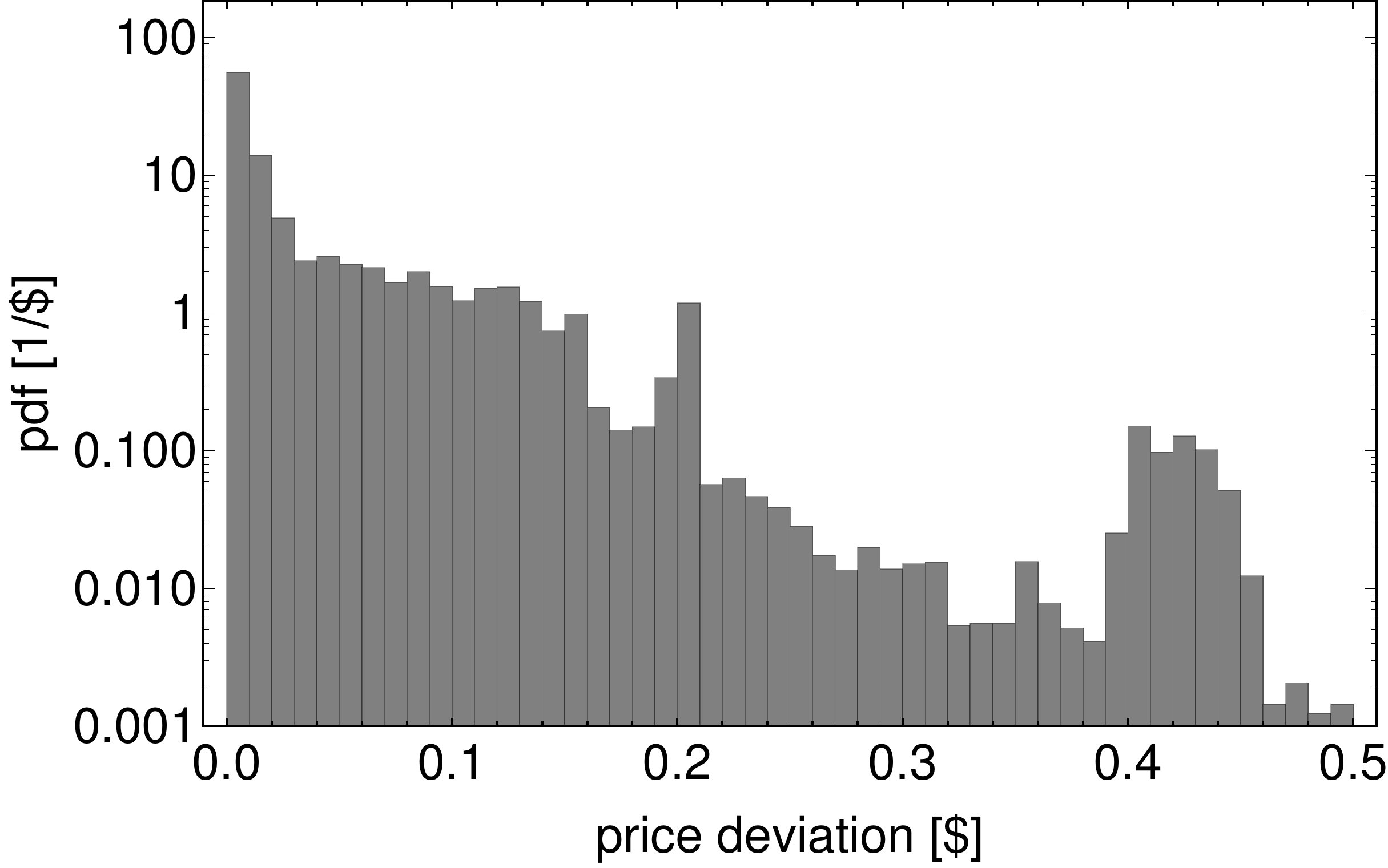}
		\label{fig:rp_sprd_hist_AAPL}
}
\subfloat[\ds{CERN}]{
		\includegraphics[width=0.45\textwidth]{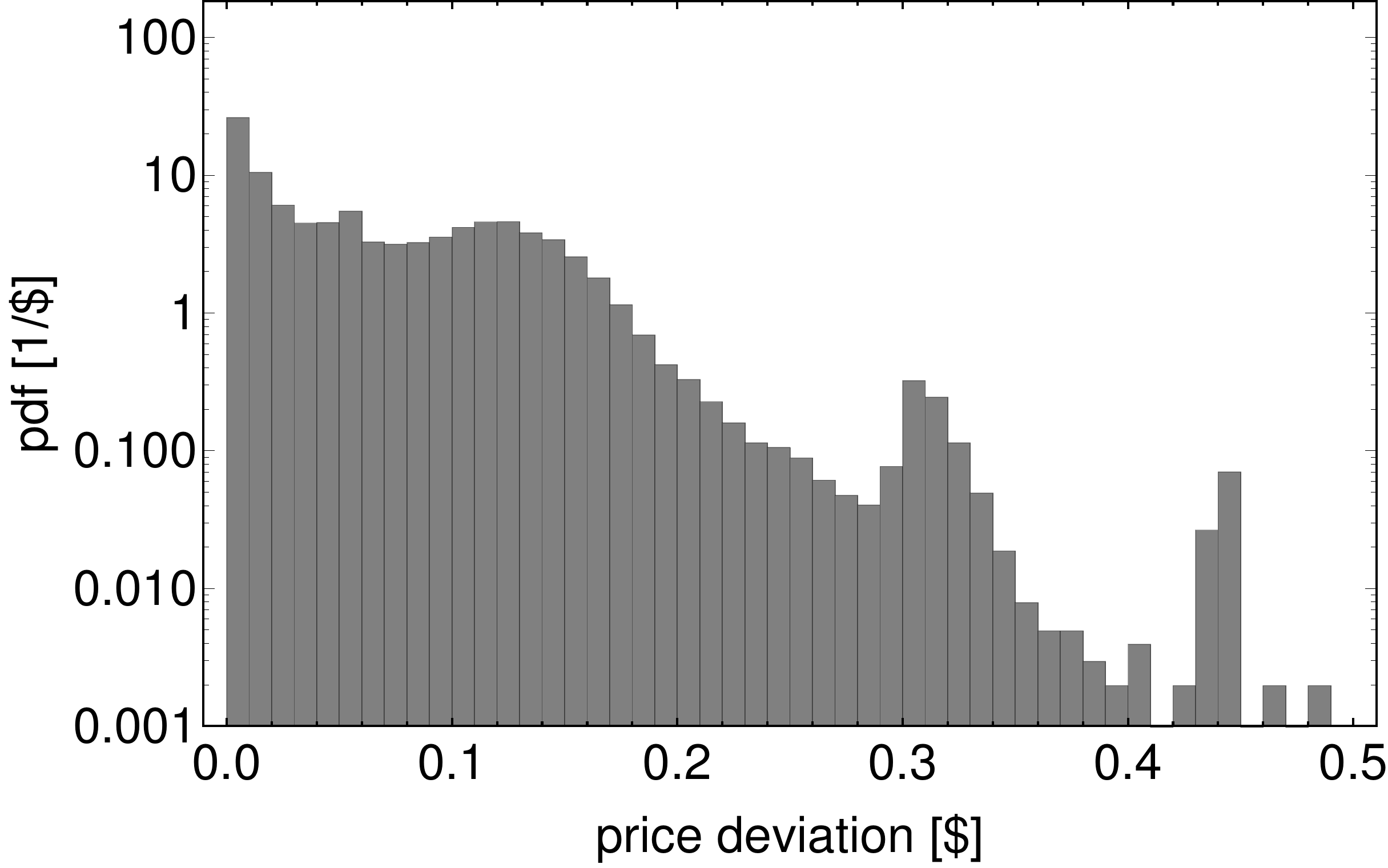}
		\label{fig:rp_sprd_hist_CERN}
}\\
\subfloat[\ds{AMGN}]{
		\includegraphics[width=0.45\textwidth]{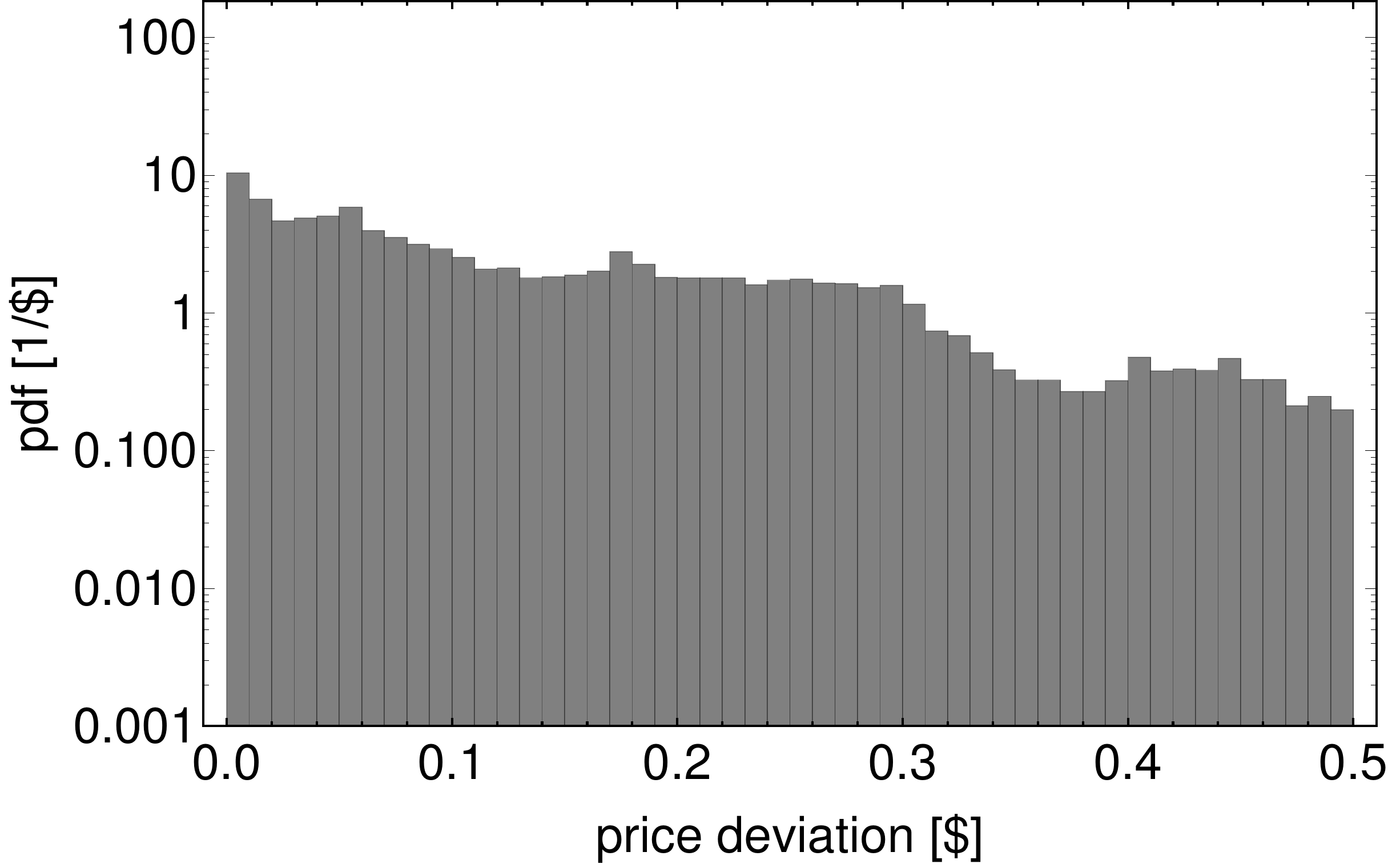}
		\label{fig:rp_sprd_hist_AMGN}
}
\subfloat[\ds{GOOG}]{
		\includegraphics[width=0.45\textwidth]{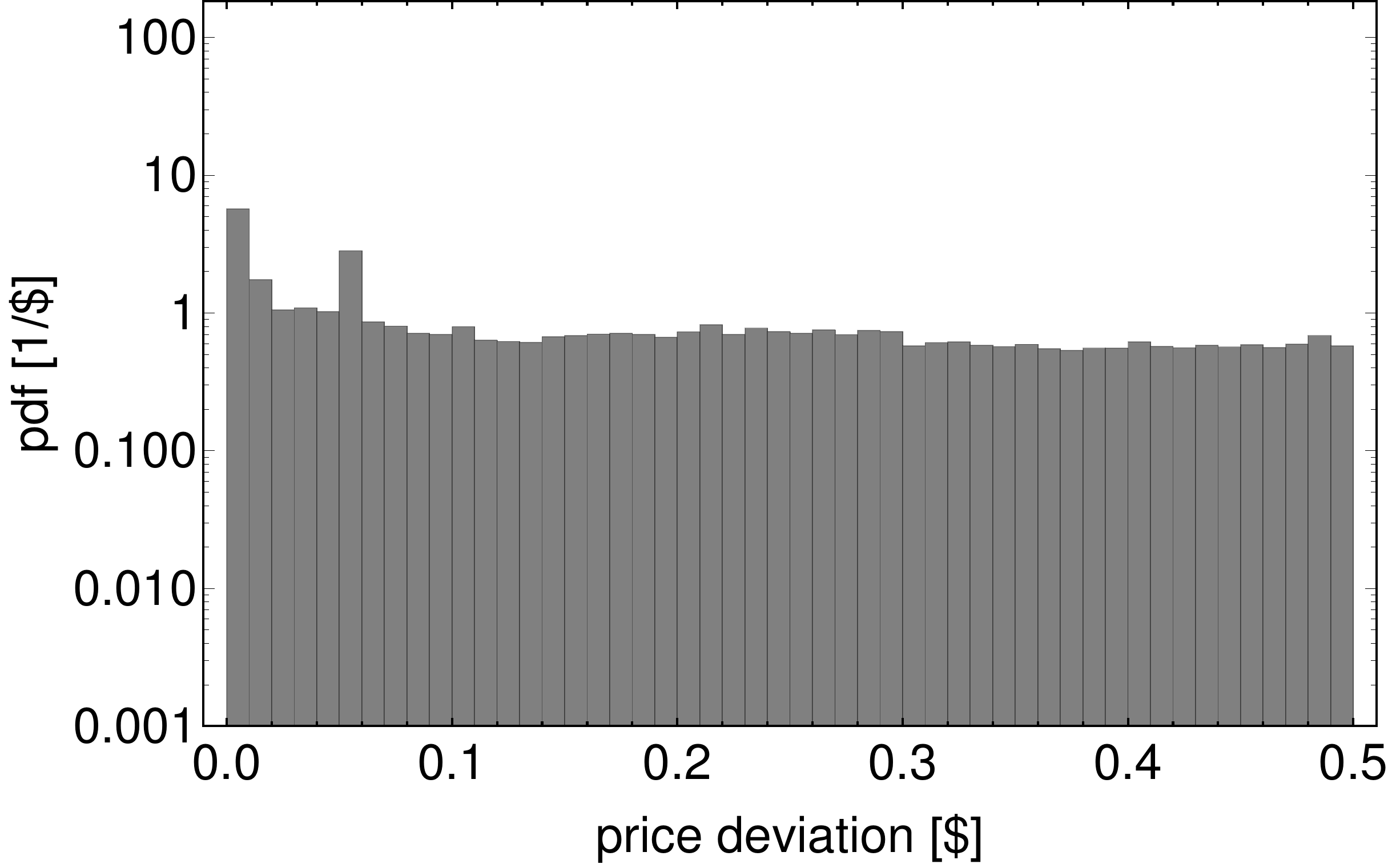}
		\label{fig:rp_sprd_hist_GOOG}
}

\caption{Distributions of absolute deviations of in-spread limit orders. The data sets shown in  \protect\subref{fig:rp_sprd_hist_AAPL} to \protect\subref{fig:rp_sprd_hist_GOOG} belong to the 
clusters C1 to C4.}
\label{fig:pd_hist}
\end{figure*}
We give an overview of the different characteristics of placements of off-spread limit orders by analyzing the relative frequency of limit orders being placed on the quotes among all 
off-spread limit orders for each stock.
It is known and consistently reported for about at least 20 years that this is where most of the incoming limit orders arrive
\cite{biais1995empirical,GouldLOBreview}.
We compute the relative frequency for buy and sell orders on each of the 480 data sets separately, see Fig. \ref{fig:rp_cluster_placement0}.
All data points scatter along the line of equality, indicating a buy-sell symmetry for this quantity. Although the clustering is done with respect to the in-spread limit 
orders, it is reflected in these relative frequencies for off-spread limit orders as well. Highest relative frequencies are seen for cluster C1 (blue circles). 
Thus especially when the spread is narrow, there is a high probability for limit orders
to be placed on the quotes. For these stocks, the order flow dynamics is essentially confined to the price levels around the quotes. 
At small relative frequencies, the stocks of cluster C4 are found (red triangles). Apparently, when the spread is broad, limit orders are not so predominantly placed on the quotes. 
In Sec.~\ref{sec:market_orders}, we show how this relates to the market order impacts. 
Once more, the two intermediate clusters C2 (yellow squares) and C3 (green diamonds) contain those stocks for which the relative frequencies are on an intermediate level. 

Since we are also interested in the distributions where limit orders are placed with respect to the quotes, we have to choose the right observable. We employ absolute deviations $\tilde d=|p-p_q|$, but relative deviations, obtained by additionally dividing by the quote price, may be used as well.
In the literature, sometimes absolute deviations \cite{Bouchaud02} and sometimes relative deviations \cite{Mike2005} are considered.
Both observables have their advantages and disadvantages. 
Especially close to the quotes, a comparison between different data sets is most suited in terms of absolute deviations $\tilde d$.
This is so, because the tick size is constant for all data sets under investigation. On the other hand, the (pseudo)discretization 
for relative deviations is different for each stock as a result of division by the quote prices, which vary by two orders of magnitude along our 
data sets and are not constant themselves within one data set, either. 
As our previous results suggest that limit orders are placed dominantly on and presumably close to the quotes, we prefer to use $\tilde d$.

Figure \ref{fig:pd_hist} displays the distributions of absolute price deviations $\tilde d$ for four data sets. Three of those are the same ones for which we showed the relative prices of in-spread limit orders. We primarily discuss differences between \ds{AAPL} and \ds{GOOG}. The role of the first bin was already discussed. It is equal to (up to a normalization factor of 100) the relative frequency for limit orders being placed on the quotes, roughly $56\%$ for \ds{AAPL} and decreases for higher cluster numbers. For \ds{GOOG} of cluster C4 we have a relative frequency of around $6\%$.
The level right next to the quotes contains roughly $14\%$ of all limit orders for \ds{AAPL} and around $2\%$ of all limit orders for \ds{GOOG}. For \ds{AAPL}, $75\%$ of all orders are found within the first three levels, while less than $9\%$ of all orders are found on these three levels for \ds{GOOG}. For the latter, the distribution of absolute price deviations decreases very slowly.
Hence, for the stocks of cluster C1, limit order arrival close to the quotes is dominant, while the depth of the order book appears to be more important for the stocks of cluster C4. Cluster C2 and C3 again feature a transition between the two extreme scenarios. 

In previous studies, power-law behavior was reported quite consistently across different markets and times studied \cite{Potters02,GouldLOBreview}. This could still be the case for stocks with large spread, especially in cluster C4. Fitting a power law to our data does not provide convincing results, maybe due to the short time window of five days. For stocks with small spreads, especially in cluster C1, the quotes and their first few neighboring levels are so dominant, that discretization effects might be too dominant for a universal power law behavior. Including data of a longer time window is beyond the scope of this study. Therefore, the characterization of off-spread limit order prices with appropriate fitting functions is left for future work. 


\subsection{Market Orders}\label{sec:market_orders}

\begin{figure}[htbp]
\centering
\includegraphics[width=0.45\textwidth]{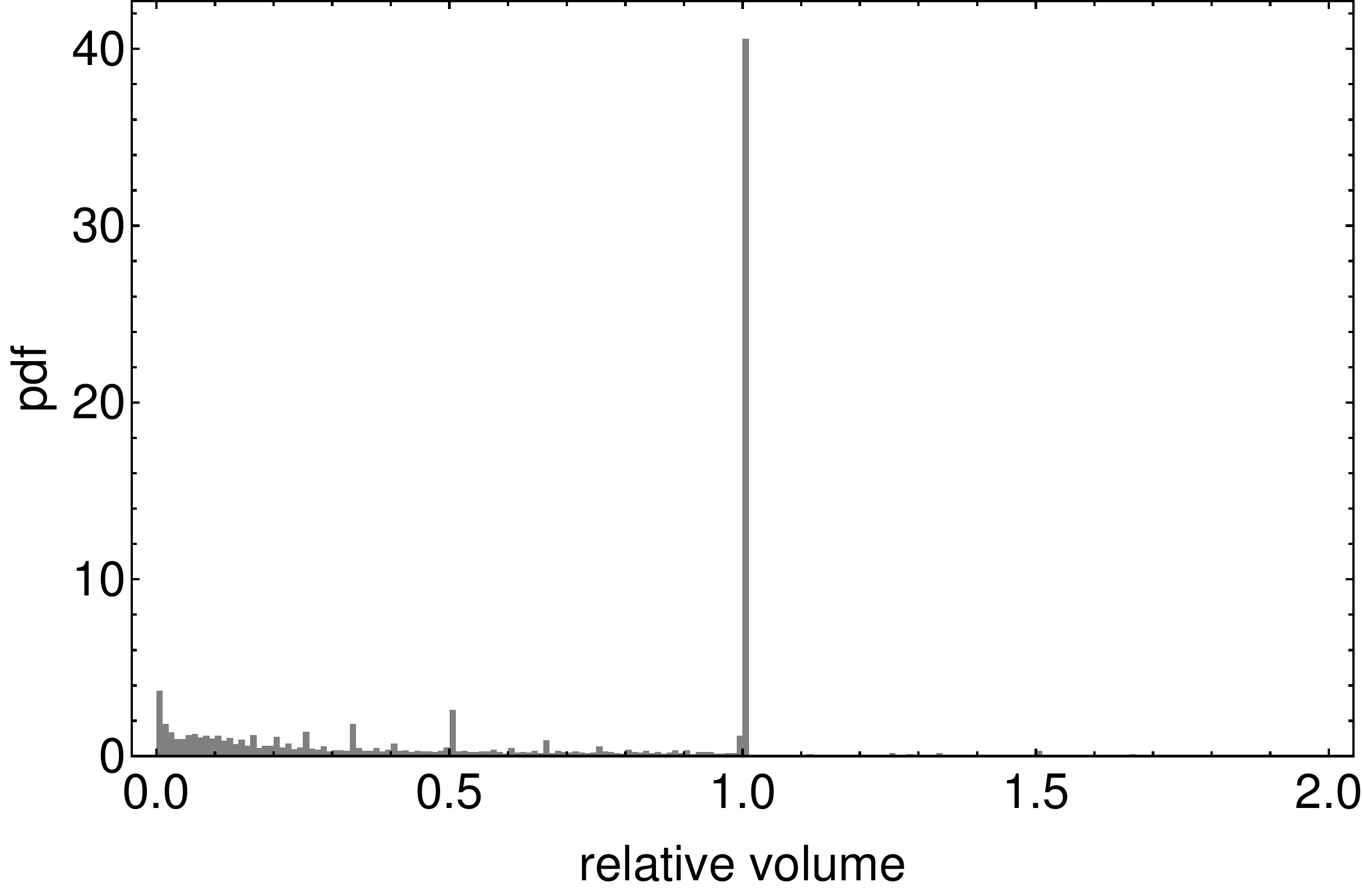}
\caption{Distributions of relative market order volumes for \ds{20160307\_AAPL}}	
\label{fig:mo_relvol_AAPL}
\end{figure}
We proceed as in \cite{Farmer2004}, where gaps in the order book were found to generate high midpoint returns. 
The mechanism is as follows: The shift of the midpoint after the arrival of a market 
order depends on the orders removed from the order book by that market order. A price 
shift occurs if the market order volume is at least as big as the volume on the 
corresponding quote.
As market order volumes are usually chosen in a way such that only the volume on the quotes 
is executed, the price shift, and therefore 
the return, depends on the distance between the quote and the next occupied level behind 
the quote. We call the distance between occupied price levels in the order book 
\textit{gaps}. The larger the gap behind the quotes, the larger the impact of a market 
order. The correspondence between the gap structure and the returns was not only found 
empirically, but was also reproduced by an agent based model \cite{Schmitt2012}. We will test how these mechanisms apply to our data.

We define the impact of market orders as
\begin{align}
r=\frac{p_{a}-p_p}{p_p},
\label{eq:mo_return}
\end{align}
where $p_p$ is the midpoint price immediately prior to the market order, and $p_a$ is 
the midpoint price
immediately after the market order.
The authors of \cite{Farmer2004} studied data from the London Stock Exchange between 1999 and 2002 and found the volumes of market orders and volumes on the 
corresponding quotes to be correlated with a correlation coefficient of $0.86$. However, this is only true if only those market orders are taken into account that change the 
midpoint. Otherwise, the correlation coefficient drops below $0.01$. This is due to a majority of market orders having a volume below the quote volume.

\begin{figure*}[htbp]
\centering
\subfloat[\ds{AAPL}]{
		\includegraphics[width=0.45\textwidth]{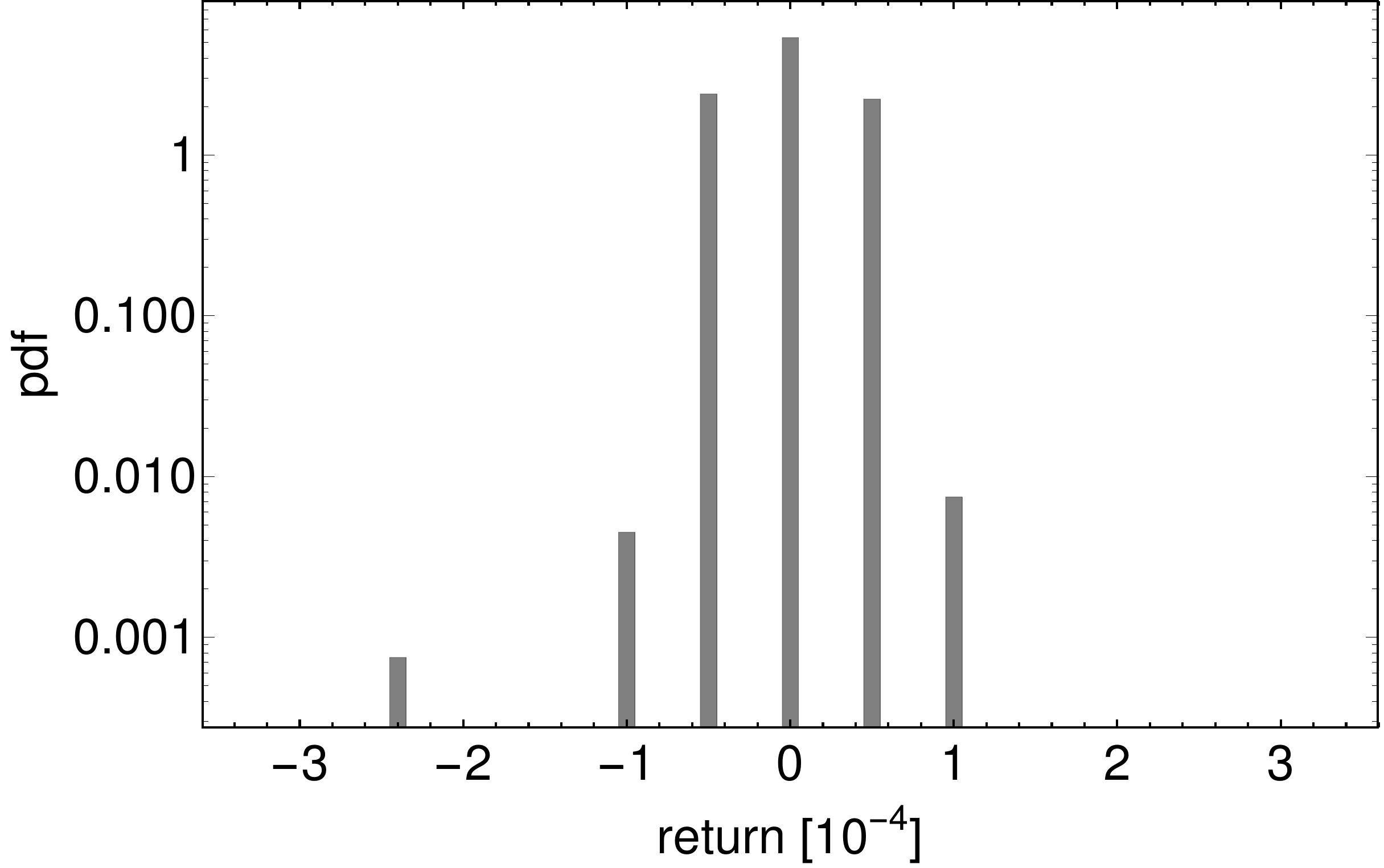}
		\label{fig:mo_imp_AAPL}
}
\subfloat[\ds{CERN}]{
		\includegraphics[width=0.45\textwidth]{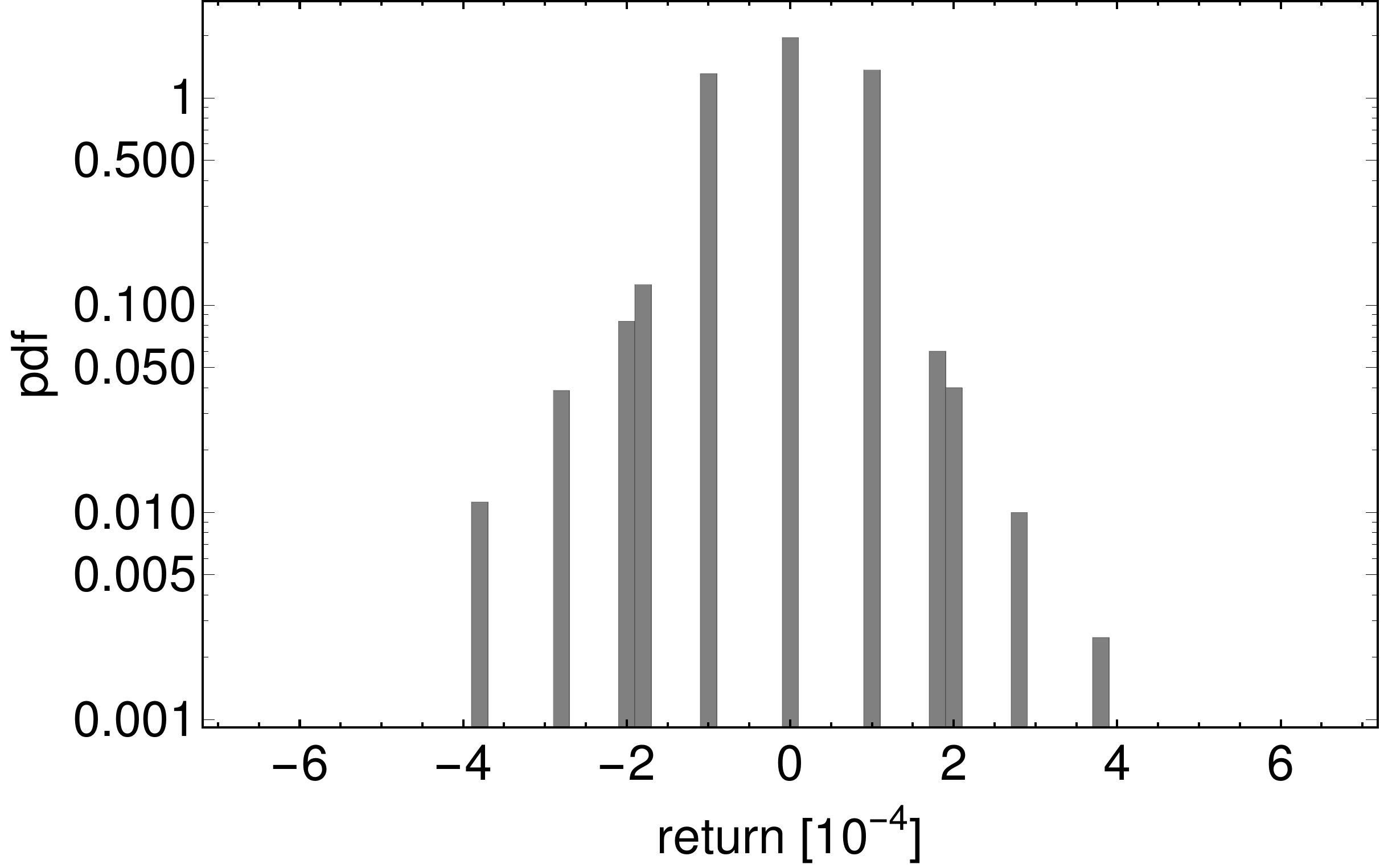}
		\label{fig:mo_imp_AMGN}
}\\
\subfloat[\ds{AMGN}]{
		\includegraphics[width=0.45\textwidth]{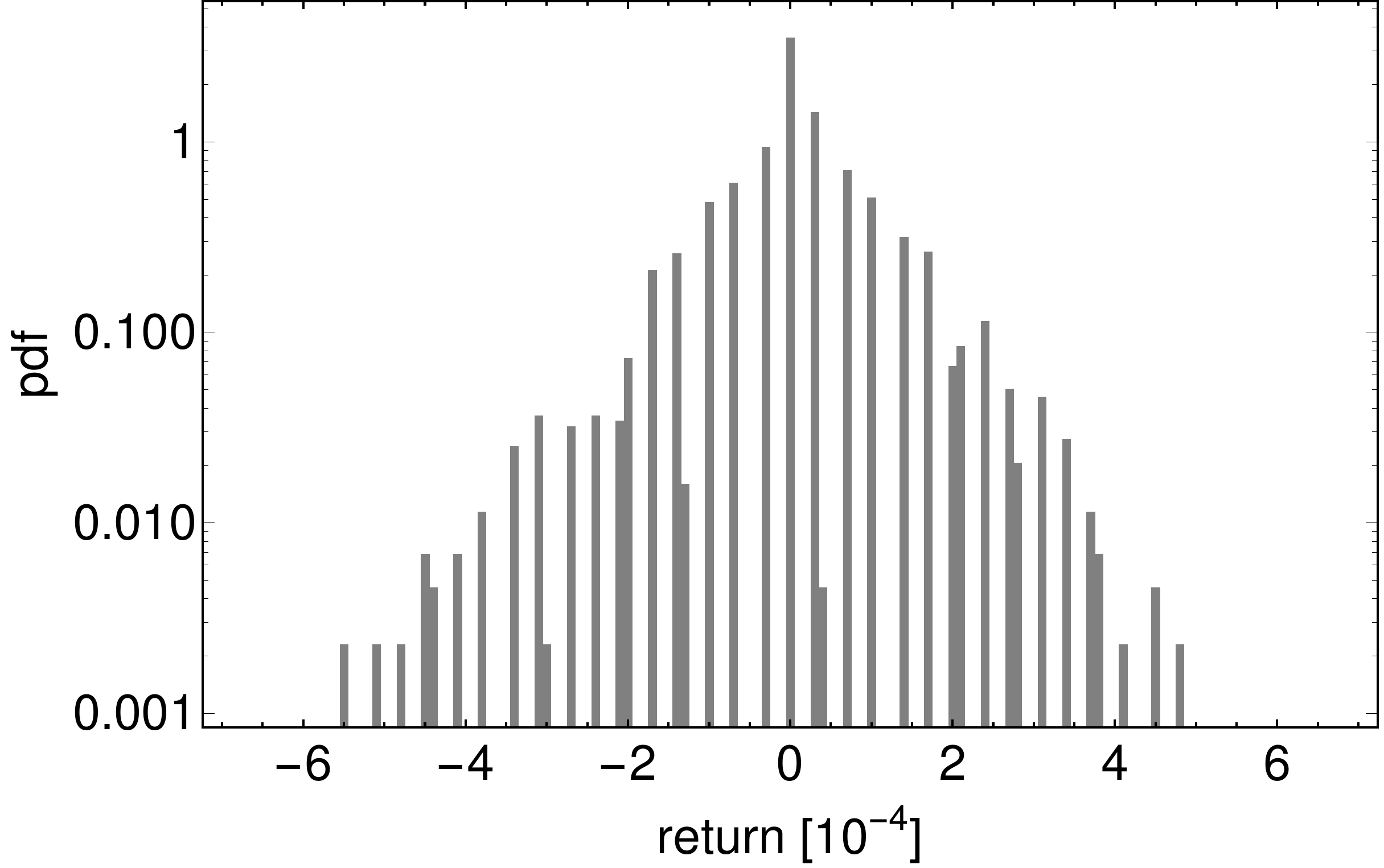}
		\label{fig:mo_imp_CERN}
}
\subfloat[\ds{GOOG}]{
		\includegraphics[width=0.45\textwidth]{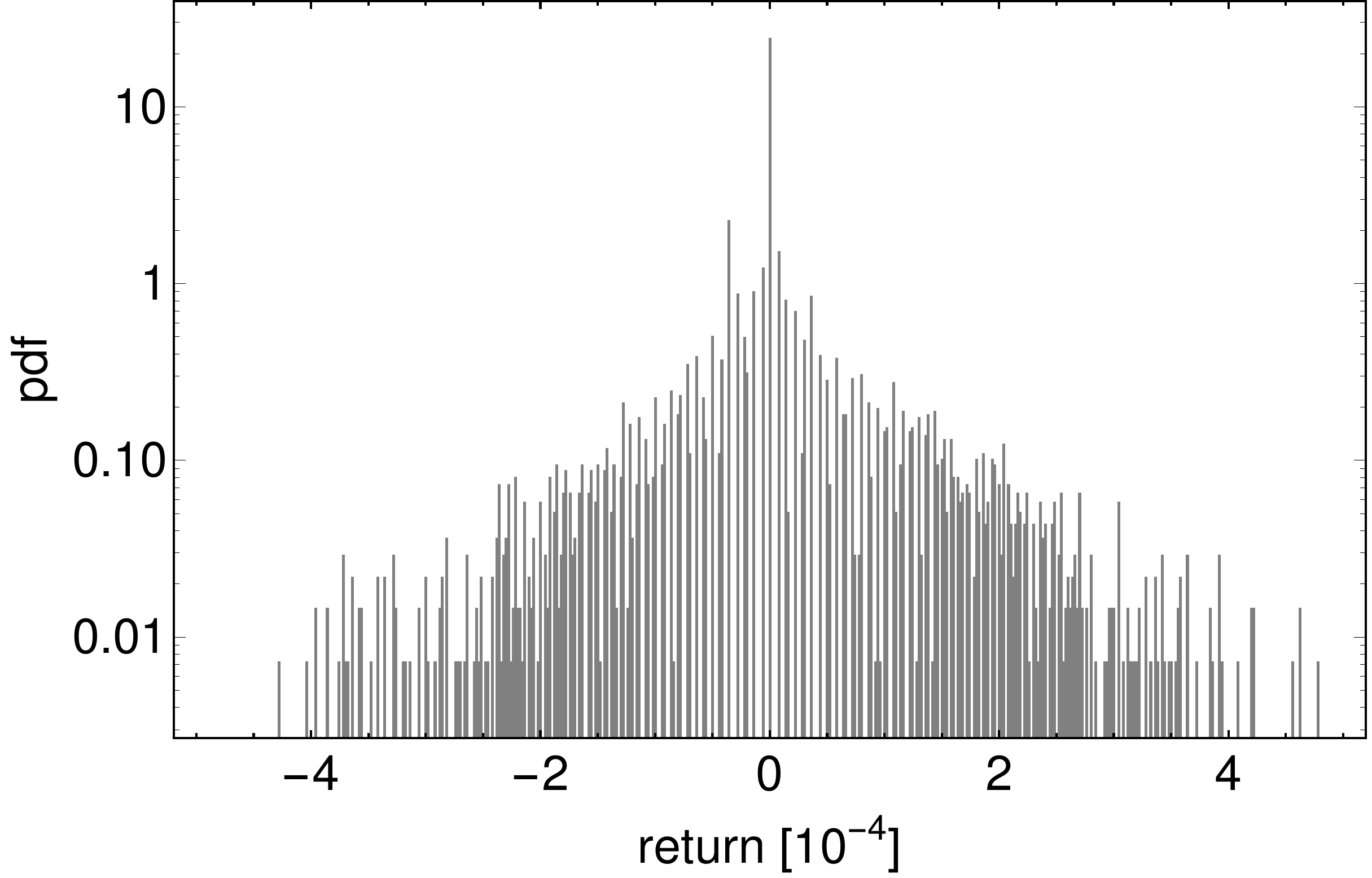}
		\label{fig:mo_imp_GOOG}
}
\caption{Distributions of returns generated by individual market orders for four stocks. The data sets shown in (a) to (d) belong to the clusters C1 to C4.}
\label{fig:MO_imp}
\end{figure*}
We find an average correlation per data set of $0.36$ when taking into account all 
market orders, regardless of having a non-zero impact or not. We consider that a consequence resulting from a higher amount 
of market orders being adjusted to have a volume equal to the quote volume, in contrast to \cite{Farmer2004}. To understand that correlation in detail, we analyze the distribution of the relative volumes of market orders.
These are defined as the market orders volumes divided by the quote volume from which liquidity is removed as the market order 
is executed. Thus, a relative volume smaller than one corresponds to a market order that does not change the midpoint, a relative volume equal to one corresponds to a market order that causes a return determined by the gap behind the quotes and a relative volume greater than one corresponds to an impact greater or equal to the gap behind the quote.

Figure \ref{fig:mo_relvol_AAPL} shows such a distribution for the data set \ds{20160307\_AAPL}. The  market order volumes match the quote volume in roughly $40.5\%$ of all cases. 
Relative volumes smaller than one have a probability of $53.1\%$. Hence, relative volumes larger than one have a probability of $6.4\%$. These results do not change 
considerably for other stocks in our data. 
This has two important implications for returns created by single market orders: 
First, we expect a high probability of around $53.1\%$ of market orders to result in zero return. Second, we note that most of 
the non-zero returns indeed follow from the gap behind the quotes. 
The strong correspondence between quote volumes and market order volumes indicates that our market order reconstruction works quite accurately.

Figure \ref{fig:MO_imp} shows the distributions of empirical returns according to Eq.~(\ref{eq:mo_return}) for our data sets.
First, we observe a pseudo-discretization for all four data sets.
The mere discretization is an effect of the tick size of $0.01\$$.
The pseudo-discretization is due to the changing midpoint, by which we 
divide when computing the returns. The authors of \cite{munnix2010impact} deal with this 
phenomenon in great detail. We choose our bin size such that the discretization is not blurred.
The number of return levels that show up reflect the sparseness of the orderbook.
We again find qualitative differences with respect to the clusters we identified in Sec.~\ref{sec:in_spread}.
For \ds{AAPL}, the order book is densely filled around the quotes, thus preventing large 
price jumps of more than a few ticks. Apparently, the more sparse price level occupation, we hypothesized to be present 
for stocks in cluster C4 like \ds{GOOG}, results in a broader return distribution. This distribution is so broad that tick size effects play a smaller 
role. Qualitatively, this fits to the results in \cite{Farmer2004}.
Due to the dense level occupation, the individual impacts tend to shrink as we move to smaller cluster numbers.

\section{Summary and Outlook}\label{sec:summary}

We analyzed order flow data of $96$ NASDAQ stocks in early 2016 and found 
remarkable qualitative differences in the order flow between different stocks.
We focused on where orders are placed in the order book, with three qualitatively
different placement positions for orders: in-spread, off-spread or
crossing the spread. The latter we referred to as effective market orders. 
We found that limit orders were submitted much more frequently than market orders, and off-spread limit orders were much more frequent than in-spread limit orders.
However, most limit orders were deleted rather than traded. 
For some stocks the small relative frequencies of in-spread limit orders resulted because the 
spread usually was equal to the tick size, out-ruling in-spread limit orders. 
Further we found that when the spread tended to be narrow, off-spread limit orders tended to be placed 
close to or on the quotes. When spreads tended to be wider, less off-spread limit orders arrived  near the quotes.
Market order volumes were typically adjusted to the quote volume. Thus, the impact of individual market orders depended on the orderbook structure, \textit{i.e.} the impact of a single market order resulted from the gap behind the spread.
When there was a high occupation density, individual market order impacts were typically reduced to the tick size (if there was any impact at all), while when the occupation density was small, impacts tended to be higher. 

With help of the relational $k$-means clustering algorithm and based on the distributions of relative prices of in-spread limit orders, we were able to grasp the diversity of order flow for different stocks. 
The diversity of the relative price distributions of in-spread limit orders could be traced back to the conditions set up by the spread. 
The different observables we took into account portrayed a consistent picture of the diversity: There were stocks with a high 
activity in terms of order arrival, where the competition of strategies led to a narrow spread and densely occupied adjacent 
levels in the orderbook. Therefore, the impact of individual orders was often constrained to the tick size. At the other end of 
the spectrum, spreads were typically large, order placement was not that 
strongly limited to the levels on and right next to the quotes, and individual impacts of market orders varied more strongly, as a 
result of a sparse occupation of price levels next to the quotes. 

Future models, such as agent based models, have to account for that diversity. It may be doubted that one model with a fixed set of mechanisms might be able to reproduce stylized facts for the different kind of dynamics we observed. 
Finally, the diversity within a fixed time window and within one market suggests that averages over a whole market should be considered carefully.

\bibliography{bibli}

\begin{thebibliography}{38}%
\makeatletter
\providecommand \@ifxundefined [1]{%
 \@ifx{#1\undefined}
}%
\providecommand \@ifnum [1]{%
 \ifnum #1\expandafter \@firstoftwo
 \else \expandafter \@secondoftwo
 \fi
}%
\providecommand \@ifx [1]{%
 \ifx #1\expandafter \@firstoftwo
 \else \expandafter \@secondoftwo
 \fi
}%
\providecommand \natexlab [1]{#1}%
\providecommand \enquote  [1]{``#1''}%
\providecommand \bibnamefont  [1]{#1}%
\providecommand \bibfnamefont [1]{#1}%
\providecommand \citenamefont [1]{#1}%
\providecommand \href@noop [0]{\@secondoftwo}%
\providecommand \href [0]{\begingroup \@sanitize@url \@href}%
\providecommand \@href[1]{\@@startlink{#1}\@@href}%
\providecommand \@@href[1]{\endgroup#1\@@endlink}%
\providecommand \@sanitize@url [0]{\catcode `\\12\catcode `\$12\catcode
  `\&12\catcode `\#12\catcode `\^12\catcode `\_12\catcode `\%12\relax}%
\providecommand \@@startlink[1]{}%
\providecommand \@@endlink[0]{}%
\providecommand \url  [0]{\begingroup\@sanitize@url \@url }%
\providecommand \@url [1]{\endgroup\@href {#1}{\urlprefix }}%
\providecommand \urlprefix  [0]{URL }%
\providecommand \Eprint [0]{\href }%
\providecommand \doibase [0]{http://dx.doi.org/}%
\providecommand \selectlanguage [0]{\@gobble}%
\providecommand \bibinfo  [0]{\@secondoftwo}%
\providecommand \bibfield  [0]{\@secondoftwo}%
\providecommand \translation [1]{[#1]}%
\providecommand \BibitemOpen [0]{}%
\providecommand \bibitemStop [0]{}%
\providecommand \bibitemNoStop [0]{.\EOS\space}%
\providecommand \EOS [0]{\spacefactor3000\relax}%
\providecommand \BibitemShut  [1]{\csname bibitem#1\endcsname}%
\let\auto@bib@innerbib\@empty
\bibitem [{\citenamefont {Farmer}\ and\ \citenamefont
  {Foley}(2009)}]{farmer2009economy}%
  \BibitemOpen
  \bibfield  {author} {\bibinfo {author} {\bibfnamefont {J~Doyne}\ \bibnamefont
  {Farmer}}\ and\ \bibinfo {author} {\bibfnamefont {Duncan}\ \bibnamefont
  {Foley}},\ }\bibfield  {title} {\enquote {\bibinfo {title} {The economy needs
  agent-based modelling},}\ }\href@noop {} {\bibfield  {journal} {\bibinfo
  {journal} {Nature}\ }\textbf {\bibinfo {volume} {460}},\ \bibinfo {pages}
  {685--686} (\bibinfo {year} {2009})}\BibitemShut {NoStop}%
\bibitem [{\citenamefont {Lee}\ \emph {et~al.}(2013)\citenamefont {Lee},
  \citenamefont {Eom},\ and\ \citenamefont {Park}}]{lee2013microstructure}%
  \BibitemOpen
  \bibfield  {author} {\bibinfo {author} {\bibfnamefont {Eun~Jung}\
  \bibnamefont {Lee}}, \bibinfo {author} {\bibfnamefont {Kyong~Shik}\
  \bibnamefont {Eom}}, \ and\ \bibinfo {author} {\bibfnamefont {Kyung~Suh}\
  \bibnamefont {Park}},\ }\bibfield  {title} {\enquote {\bibinfo {title}
  {Microstructure-based manipulation: Strategic behavior and performance of
  spoofing traders},}\ }\href@noop {} {\bibfield  {journal} {\bibinfo
  {journal} {Journal of Financial Markets}\ }\textbf {\bibinfo {volume} {16}},\
  \bibinfo {pages} {227--252} (\bibinfo {year} {2013})}\BibitemShut {NoStop}%
\bibitem [{\citenamefont {Kirilenko}\ \emph {et~al.}(2015)\citenamefont
  {Kirilenko}, \citenamefont {Kyle}, \citenamefont {Samadi},\ and\
  \citenamefont {Tuzun}}]{kirilenko2015flash}%
  \BibitemOpen
  \bibfield  {author} {\bibinfo {author} {\bibfnamefont {Andrei~A}\
  \bibnamefont {Kirilenko}}, \bibinfo {author} {\bibfnamefont {Albert~S}\
  \bibnamefont {Kyle}}, \bibinfo {author} {\bibfnamefont {Mehrdad}\
  \bibnamefont {Samadi}}, \ and\ \bibinfo {author} {\bibfnamefont {Tugkan}\
  \bibnamefont {Tuzun}},\ }\bibfield  {title} {\enquote {\bibinfo {title} {The
  flash crash: The impact of high frequency trading on an electronic market},}\
  }\href@noop {} {\bibfield  {journal} {\bibinfo  {journal} {Available at SSRN
  1686004}\ } (\bibinfo {year} {2015})}\BibitemShut {NoStop}%
\bibitem [{\citenamefont {Brogaard}(2010)}]{brogaard2010high}%
  \BibitemOpen
  \bibfield  {author} {\bibinfo {author} {\bibfnamefont {Jonathan}\
  \bibnamefont {Brogaard}},\ }\bibfield  {title} {\enquote {\bibinfo {title}
  {High frequency trading and its impact on market quality},}\ }\href@noop {}
  {\bibfield  {journal} {\bibinfo  {journal} {Northwestern University Kellogg
  School of Management Working Paper}\ }\textbf {\bibinfo {volume} {66}}
  (\bibinfo {year} {2010})}\BibitemShut {NoStop}%
\bibitem [{\citenamefont {Patzelt}\ and\ \citenamefont
  {Pawelzik}(2013)}]{patzelt2013inherent}%
  \BibitemOpen
  \bibfield  {author} {\bibinfo {author} {\bibfnamefont {Felix}\ \bibnamefont
  {Patzelt}}\ and\ \bibinfo {author} {\bibfnamefont {Klaus}\ \bibnamefont
  {Pawelzik}},\ }\bibfield  {title} {\enquote {\bibinfo {title} {An inherent
  instability of efficient markets},}\ }\href@noop {} {\bibfield  {journal}
  {\bibinfo  {journal} {Scientific reports}\ }\textbf {\bibinfo {volume} {3}}
  (\bibinfo {year} {2013})}\BibitemShut {NoStop}%
\bibitem [{\citenamefont {Meudt}\ \emph {et~al.}(2016)\citenamefont {Meudt},
  \citenamefont {Schmitt}, \citenamefont {Sch{\"a}fer},\ and\ \citenamefont
  {Guhr}}]{meudt2016equilibrium}%
  \BibitemOpen
  \bibfield  {author} {\bibinfo {author} {\bibfnamefont {Frederik}\
  \bibnamefont {Meudt}}, \bibinfo {author} {\bibfnamefont {Thilo~A}\
  \bibnamefont {Schmitt}}, \bibinfo {author} {\bibfnamefont {Rudi}\
  \bibnamefont {Sch{\"a}fer}}, \ and\ \bibinfo {author} {\bibfnamefont
  {Thomas}\ \bibnamefont {Guhr}},\ }\bibfield  {title} {\enquote {\bibinfo
  {title} {Equilibrium pricing in an order book environment: Case study for a
  spin model},}\ }\href@noop {} {\bibfield  {journal} {\bibinfo  {journal}
  {Physica A: Statistical Mechanics and its Applications}\ }\textbf {\bibinfo
  {volume} {453}},\ \bibinfo {pages} {228--235} (\bibinfo {year}
  {2016})}\BibitemShut {NoStop}%
\bibitem [{\citenamefont {Krause}\ and\ \citenamefont
  {Bornholdt}(2013)}]{krause2013spin}%
  \BibitemOpen
  \bibfield  {author} {\bibinfo {author} {\bibfnamefont {Sebastian~M}\
  \bibnamefont {Krause}}\ and\ \bibinfo {author} {\bibfnamefont {Stefan}\
  \bibnamefont {Bornholdt}},\ }\bibfield  {title} {\enquote {\bibinfo {title}
  {Spin models as microfoundation of macroscopic market models},}\ }\href@noop
  {} {\bibfield  {journal} {\bibinfo  {journal} {Physica A: Statistical
  Mechanics and its Applications}\ }\textbf {\bibinfo {volume} {392}},\
  \bibinfo {pages} {4048--4054} (\bibinfo {year} {2013})}\BibitemShut {NoStop}%
\bibitem [{\citenamefont {Cont}(2001)}]{cont2001empirical}%
  \BibitemOpen
  \bibfield  {author} {\bibinfo {author} {\bibfnamefont {Rama}\ \bibnamefont
  {Cont}},\ }\bibfield  {title} {\enquote {\bibinfo {title} {{Empirical
  properties of asset returns: stylized facts and statistical issues}},}\
  }\href@noop {} {\  (\bibinfo {year} {2001})}\BibitemShut {NoStop}%
\bibitem [{\citenamefont {Gould}\ \emph {et~al.}(2013)\citenamefont {Gould},
  \citenamefont {Porter}, \citenamefont {Williams}, \citenamefont {McDonald},
  \citenamefont {Fenn},\ and\ \citenamefont {Howison}}]{GouldLOBreview}%
  \BibitemOpen
  \bibfield  {author} {\bibinfo {author} {\bibfnamefont {Martin~D}\
  \bibnamefont {Gould}}, \bibinfo {author} {\bibfnamefont {Mason~A}\
  \bibnamefont {Porter}}, \bibinfo {author} {\bibfnamefont {Stacy}\
  \bibnamefont {Williams}}, \bibinfo {author} {\bibfnamefont {Mark}\
  \bibnamefont {McDonald}}, \bibinfo {author} {\bibfnamefont {Daniel~J}\
  \bibnamefont {Fenn}}, \ and\ \bibinfo {author} {\bibfnamefont {Sam~D}\
  \bibnamefont {Howison}},\ }\bibfield  {title} {\enquote {\bibinfo {title}
  {{Limit order books}},}\ }\href {?} {\bibfield  {journal} {\bibinfo
  {journal} {Quantitative Finance}\ }\textbf {\bibinfo {volume} {13}},\
  \bibinfo {pages} {1709--1742} (\bibinfo {year} {2013})}\BibitemShut {NoStop}%
\bibitem [{\citenamefont {{Chakrabarti, Anindya S.}}\ and\ \citenamefont
  {{Lahkar, Ratul}}(2016)}]{chakrabarti2016absence}%
  \BibitemOpen
  \bibfield  {author} {\bibinfo {author} {\bibnamefont {{Chakrabarti, Anindya
  S.}}}\ and\ \bibinfo {author} {\bibnamefont {{Lahkar, Ratul}}},\ }\bibfield
  {title} {\enquote {\bibinfo {title} {Absence of economic and social
  constants},}\ }\href {\doibase 10.1140/epjst/e2016-60176-3} {\bibfield
  {journal} {\bibinfo  {journal} {Eur. Phys. J. Special Topics}\ }\textbf
  {\bibinfo {volume} {225}} (\bibinfo {year} {2016}),\
  10.1140/epjst/e2016-60176-3}\BibitemShut {NoStop}%
\bibitem [{\citenamefont {Potters}\ and\ \citenamefont
  {Bouchaud}(2003)}]{Potters02}%
  \BibitemOpen
  \bibfield  {author} {\bibinfo {author} {\bibfnamefont {Marc}\ \bibnamefont
  {Potters}}\ and\ \bibinfo {author} {\bibfnamefont {Jean-Philippe}\
  \bibnamefont {Bouchaud}},\ }\bibfield  {title} {\enquote {\bibinfo {title}
  {{More statistical properties of order books and price impact}},}\ }\href {?}
  {\bibfield  {journal} {\bibinfo  {journal} {Physica A: Statistical Mechanics
  and its Applications}\ }\textbf {\bibinfo {volume} {324}},\ \bibinfo {pages}
  {133--140} (\bibinfo {year} {2003})}\BibitemShut {NoStop}%
\bibitem [{\citenamefont {Gareche}\ \emph {et~al.}(2013)\citenamefont
  {Gareche}, \citenamefont {Disdier}, \citenamefont {Kockelkoren},\ and\
  \citenamefont {Bouchaud}}]{gareche2013fokker}%
  \BibitemOpen
  \bibfield  {author} {\bibinfo {author} {\bibfnamefont {A}~\bibnamefont
  {Gareche}}, \bibinfo {author} {\bibfnamefont {G}~\bibnamefont {Disdier}},
  \bibinfo {author} {\bibfnamefont {J}~\bibnamefont {Kockelkoren}}, \ and\
  \bibinfo {author} {\bibfnamefont {J-P}\ \bibnamefont {Bouchaud}},\ }\bibfield
   {title} {\enquote {\bibinfo {title} {Fokker-planck description for the queue
  dynamics of large tick stocks},}\ }\href@noop {} {\bibfield  {journal}
  {\bibinfo  {journal} {Physical Review E}\ }\textbf {\bibinfo {volume} {88}},\
  \bibinfo {pages} {032809} (\bibinfo {year} {2013})}\BibitemShut {NoStop}%
\bibitem [{\citenamefont {Wang}\ \emph {et~al.}(2016)\citenamefont {Wang},
  \citenamefont {Sch{\"a}fer},\ and\ \citenamefont {Guhr}}]{Wang2016}%
  \BibitemOpen
  \bibfield  {author} {\bibinfo {author} {\bibfnamefont {Shanshan}\
  \bibnamefont {Wang}}, \bibinfo {author} {\bibfnamefont {Rudi}\ \bibnamefont
  {Sch{\"a}fer}}, \ and\ \bibinfo {author} {\bibfnamefont {Thomas}\
  \bibnamefont {Guhr}},\ }\bibfield  {title} {\enquote {\bibinfo {title}
  {{Average cross-responses in correlated financial market}},}\ }\href {?}
  {\bibfield  {journal} {\bibinfo  {journal} {arXiv preprint arXiv:1603.01586}\
  } (\bibinfo {year} {2016})}\BibitemShut {NoStop}%
\bibitem [{\citenamefont {Hasbrouck}\ and\ \citenamefont
  {Seppi}(2001)}]{hasbrouck2001common}%
  \BibitemOpen
  \bibfield  {author} {\bibinfo {author} {\bibfnamefont {Joel}\ \bibnamefont
  {Hasbrouck}}\ and\ \bibinfo {author} {\bibfnamefont {Duane~J}\ \bibnamefont
  {Seppi}},\ }\bibfield  {title} {\enquote {\bibinfo {title} {Common factors in
  prices, order flows, and liquidity},}\ }\href@noop {} {\bibfield  {journal}
  {\bibinfo  {journal} {Journal of financial Economics}\ }\textbf {\bibinfo
  {volume} {59}},\ \bibinfo {pages} {383--411} (\bibinfo {year}
  {2001})}\BibitemShut {NoStop}%
\bibitem [{\citenamefont {Boulatov}\ \emph {et~al.}(2013)\citenamefont
  {Boulatov}, \citenamefont {Hendershott},\ and\ \citenamefont
  {Livdan}}]{boulatov2013informed}%
  \BibitemOpen
  \bibfield  {author} {\bibinfo {author} {\bibfnamefont {Alex}\ \bibnamefont
  {Boulatov}}, \bibinfo {author} {\bibfnamefont {Terrence}\ \bibnamefont
  {Hendershott}}, \ and\ \bibinfo {author} {\bibfnamefont {Dmitry}\
  \bibnamefont {Livdan}},\ }\bibfield  {title} {\enquote {\bibinfo {title}
  {Informed trading and portfolio returns},}\ }\href@noop {} {\bibfield
  {journal} {\bibinfo  {journal} {The Review of Economic Studies}\ }\textbf
  {\bibinfo {volume} {80}},\ \bibinfo {pages} {35--72} (\bibinfo {year}
  {2013})}\BibitemShut {NoStop}%
\bibitem [{\citenamefont {Pasquariello}\ and\ \citenamefont
  {Vega}(2013)}]{pasquariello2013strategic}%
  \BibitemOpen
  \bibfield  {author} {\bibinfo {author} {\bibfnamefont {Paolo}\ \bibnamefont
  {Pasquariello}}\ and\ \bibinfo {author} {\bibfnamefont {Clara}\ \bibnamefont
  {Vega}},\ }\bibfield  {title} {\enquote {\bibinfo {title} {Strategic
  cross-trading in the us stock market},}\ }\href@noop {} {\bibfield  {journal}
  {\bibinfo  {journal} {Review of Finance}\ ,\ \bibinfo {pages} {rft055}}
  (\bibinfo {year} {2013})}\BibitemShut {NoStop}%
\bibitem [{\citenamefont {Chordia}\ \emph {et~al.}(2000)\citenamefont
  {Chordia}, \citenamefont {Roll},\ and\ \citenamefont
  {Subrahmanyam}}]{chordia2000commonality}%
  \BibitemOpen
  \bibfield  {author} {\bibinfo {author} {\bibfnamefont {Tarun}\ \bibnamefont
  {Chordia}}, \bibinfo {author} {\bibfnamefont {Richard}\ \bibnamefont {Roll}},
  \ and\ \bibinfo {author} {\bibfnamefont {Avanidhar}\ \bibnamefont
  {Subrahmanyam}},\ }\bibfield  {title} {\enquote {\bibinfo {title}
  {Commonality in liquidity},}\ }\href@noop {} {\bibfield  {journal} {\bibinfo
  {journal} {Journal of financial economics}\ }\textbf {\bibinfo {volume}
  {56}},\ \bibinfo {pages} {3--28} (\bibinfo {year} {2000})}\BibitemShut
  {NoStop}%
\bibitem [{\citenamefont {Stepanov}\ \emph {et~al.}(2015)\citenamefont
  {Stepanov}, \citenamefont {Rinn}, \citenamefont {Guhr}, \citenamefont
  {Peinke},\ and\ \citenamefont {Sch{\"a}fer}}]{stepanov2015stability}%
  \BibitemOpen
  \bibfield  {author} {\bibinfo {author} {\bibfnamefont {Yuriy}\ \bibnamefont
  {Stepanov}}, \bibinfo {author} {\bibfnamefont {Philip}\ \bibnamefont {Rinn}},
  \bibinfo {author} {\bibfnamefont {Thomas}\ \bibnamefont {Guhr}}, \bibinfo
  {author} {\bibfnamefont {Joachim}\ \bibnamefont {Peinke}}, \ and\ \bibinfo
  {author} {\bibfnamefont {Rudi}\ \bibnamefont {Sch{\"a}fer}},\ }\bibfield
  {title} {\enquote {\bibinfo {title} {Stability and hierarchy of
  quasi-stationary states: financial markets as an example},}\ }\href@noop {}
  {\bibfield  {journal} {\bibinfo  {journal} {Journal of Statistical Mechanics:
  Theory and Experiment}\ }\textbf {\bibinfo {volume} {2015}},\ \bibinfo
  {pages} {P08011} (\bibinfo {year} {2015})}\BibitemShut {NoStop}%
\bibitem [{\citenamefont {https://en.wikipedia.org/wiki/NASDAQ
  100}()}]{wikiNasdaq100}%
  \BibitemOpen
  \bibfield  {author} {\bibinfo {author} {\bibnamefont
  {https://en.wikipedia.org/wiki/NASDAQ 100}},\ }\href@noop {} {}\bibinfo
  {note} {Accessed: September 28, 2016}\BibitemShut {NoStop}%
\bibitem [{\citenamefont {http://tradingphysics.com/}()}]{tp}%
  \BibitemOpen
  \bibfield  {author} {\bibinfo {author} {\bibnamefont
  {http://tradingphysics.com/}},\ }\href@noop {} {}\bibinfo {note} {Accessed:
  September 14, 2016}\BibitemShut {NoStop}%
\bibitem [{\citenamefont {Huang}\ and\ \citenamefont
  {Polak}(2011)}]{huang2011lobster}%
  \BibitemOpen
  \bibfield  {author} {\bibinfo {author} {\bibfnamefont {Ruihong}\ \bibnamefont
  {Huang}}\ and\ \bibinfo {author} {\bibfnamefont {Tomas}\ \bibnamefont
  {Polak}},\ }\bibfield  {title} {\enquote {\bibinfo {title} {Lobster: Limit
  order book reconstruction system},}\ }\href@noop {} {\bibfield  {journal}
  {\bibinfo  {journal} {Available at SSRN 1977207}\ } (\bibinfo {year}
  {2011})}\BibitemShut {NoStop}%
\bibitem [{\citenamefont
  {{http://www.nasdaqtrader.com/content/Products/\\Services/Trading/OrderTypesG.pdf}}()}]{order_types}%
  \BibitemOpen
  \bibfield  {author} {\bibinfo {author} {\bibnamefont
  {{http://www.nasdaqtrader.com/content/Products/\\Services/Trading/OrderTypesG.pdf}}},\
  }\href@noop {} {}\bibinfo {note} {Accessed: November 21, 2016}\BibitemShut
  {NoStop}%
\bibitem [{\citenamefont
  {{http://howtohft.blogspot.de/2012/07/tradingphysics-historical-totalview.html}}()}]{howtohft_blog}%
  \BibitemOpen
  \bibfield  {author} {\bibinfo {author} {\bibnamefont
  {{http://howtohft.blogspot.de/2012/07/tradingphysics-historical-totalview.html}}},\
  }\href@noop {} {}\bibinfo {note} {Accessed: September 13, 2016}\BibitemShut
  {NoStop}%
\bibitem [{\citenamefont
  {{http://quant.caltech.edu/historical-stock-data.html}}()}]{caltech}%
  \BibitemOpen
  \bibfield  {author} {\bibinfo {author} {\bibnamefont
  {{http://quant.caltech.edu/historical-stock-data.html}}},\ }\href@noop {}
  {}\bibinfo {note} {Accessed: September 13, 2016}\BibitemShut {NoStop}%
\bibitem [{\citenamefont {Hautsch}\ and\ \citenamefont
  {Huang}(2011)}]{Hautsch2011-2}%
  \BibitemOpen
  \bibfield  {author} {\bibinfo {author} {\bibfnamefont {Nikolaus}\
  \bibnamefont {Hautsch}}\ and\ \bibinfo {author} {\bibfnamefont {Ruihong}\
  \bibnamefont {Huang}},\ }\bibfield  {title} {\enquote {\bibinfo {title}
  {{Limit order flow, market impact and optimal order sizes: evidence from
  NASDAQ TotalView-ITCH data}},}\ }\href {?} {\bibfield  {journal} {\bibinfo
  {journal} {Market Impact and Optimal Order Sizes: Evidence from NASDAQ
  TotalView-ITCH Data (August 22, 2011)}\ } (\bibinfo {year}
  {2011})}\BibitemShut {NoStop}%
\bibitem [{\citenamefont
  {http://www.nasdaq.com/services/homw.stm}()}]{Nasdaq_execution_time}%
  \BibitemOpen
  \bibfield  {author} {\bibinfo {author} {\bibnamefont
  {http://www.nasdaq.com/services/homw.stm}},\ }\href@noop {} {}\bibinfo {note}
  {Accessed: August 11, 2016}\BibitemShut {NoStop}%
\bibitem [{\citenamefont {Bouchaud}\ \emph {et~al.}(2002)\citenamefont
  {Bouchaud}, \citenamefont {M{\'e}zard}, \citenamefont {Potters} \emph
  {et~al.}}]{Bouchaud02}%
  \BibitemOpen
  \bibfield  {author} {\bibinfo {author} {\bibfnamefont {Jean-Philippe}\
  \bibnamefont {Bouchaud}}, \bibinfo {author} {\bibfnamefont {Marc}\
  \bibnamefont {M{\'e}zard}}, \bibinfo {author} {\bibfnamefont {Marc}\
  \bibnamefont {Potters}},  \emph {et~al.},\ }\bibfield  {title} {\enquote
  {\bibinfo {title} {{Statistical properties of stock order books: empirical
  results and models}},}\ }\href {?} {\bibfield  {journal} {\bibinfo  {journal}
  {Quantitative finance}\ }\textbf {\bibinfo {volume} {2}},\ \bibinfo {pages}
  {251--256} (\bibinfo {year} {2002})}\BibitemShut {NoStop}%
\bibitem [{\citenamefont {Szalkai}(2013)}]{Szalkai2013}%
  \BibitemOpen
  \bibfield  {author} {\bibinfo {author} {\bibfnamefont {Bal{\'a}zs}\
  \bibnamefont {Szalkai}},\ }\bibfield  {title} {\enquote {\bibinfo {title}
  {{An implementation of the relational k-means algorithm}},}\ }\href@noop {}
  {\bibfield  {journal} {\bibinfo  {journal} {arXiv preprint arXiv:1304.6899}\
  } (\bibinfo {year} {2013})}\BibitemShut {NoStop}%
\bibitem [{\citenamefont {Rousseeuw}(1987)}]{rousseeuw1987silhouettes}%
  \BibitemOpen
  \bibfield  {author} {\bibinfo {author} {\bibfnamefont {Peter~J}\ \bibnamefont
  {Rousseeuw}},\ }\bibfield  {title} {\enquote {\bibinfo {title} {Silhouettes:
  a graphical aid to the interpretation and validation of cluster analysis},}\
  }\href@noop {} {\bibfield  {journal} {\bibinfo  {journal} {Journal of
  computational and applied mathematics}\ }\textbf {\bibinfo {volume} {20}},\
  \bibinfo {pages} {53--65} (\bibinfo {year} {1987})}\BibitemShut {NoStop}%
\bibitem [{\citenamefont {Eisler}\ \emph {et~al.}(2012)\citenamefont {Eisler},
  \citenamefont {Bouchaud},\ and\ \citenamefont
  {Kockelkoren}}]{eisler2012price}%
  \BibitemOpen
  \bibfield  {author} {\bibinfo {author} {\bibfnamefont {Zoltan}\ \bibnamefont
  {Eisler}}, \bibinfo {author} {\bibfnamefont {Jean-Philippe}\ \bibnamefont
  {Bouchaud}}, \ and\ \bibinfo {author} {\bibfnamefont {Julien}\ \bibnamefont
  {Kockelkoren}},\ }\bibfield  {title} {\enquote {\bibinfo {title} {The price
  impact of order book events: market orders, limit orders and
  cancellations},}\ }\href@noop {} {\bibfield  {journal} {\bibinfo  {journal}
  {Quantitative Finance}\ }\textbf {\bibinfo {volume} {12}},\ \bibinfo {pages}
  {1395--1419} (\bibinfo {year} {2012})}\BibitemShut {NoStop}%
\bibitem [{\citenamefont {Dayri}\ and\ \citenamefont
  {Rosenbaum}(2015)}]{dayri2015large}%
  \BibitemOpen
  \bibfield  {author} {\bibinfo {author} {\bibfnamefont {Khalil}\ \bibnamefont
  {Dayri}}\ and\ \bibinfo {author} {\bibfnamefont {Mathieu}\ \bibnamefont
  {Rosenbaum}},\ }\bibfield  {title} {\enquote {\bibinfo {title} {Large tick
  assets: implicit spread and optimal tick size},}\ }\href@noop {} {\bibfield
  {journal} {\bibinfo  {journal} {Market Microstructure and Liquidity}\
  }\textbf {\bibinfo {volume} {1}},\ \bibinfo {pages} {1550003} (\bibinfo
  {year} {2015})}\BibitemShut {NoStop}%
\bibitem [{\citenamefont {Doyne~Farmer}\ \emph {et~al.}(2004)\citenamefont
  {Doyne~Farmer}, \citenamefont {Gillemot}, \citenamefont {Lillo},
  \citenamefont {Mike},\ and\ \citenamefont {Sen}}]{Farmer2004}%
  \BibitemOpen
  \bibfield  {author} {\bibinfo {author} {\bibfnamefont {J}~\bibnamefont
  {Doyne~Farmer}}, \bibinfo {author} {\bibfnamefont {Laszlo}\ \bibnamefont
  {Gillemot}}, \bibinfo {author} {\bibfnamefont {Fabrizio}\ \bibnamefont
  {Lillo}}, \bibinfo {author} {\bibfnamefont {Szabolcs}\ \bibnamefont {Mike}},
  \ and\ \bibinfo {author} {\bibfnamefont {Anindya}\ \bibnamefont {Sen}},\
  }\bibfield  {title} {\enquote {\bibinfo {title} {{What really causes large
  price changes?}}}\ }\href {?} {\bibfield  {journal} {\bibinfo  {journal}
  {Quantitative finance}\ }\textbf {\bibinfo {volume} {4}},\ \bibinfo {pages}
  {383--397} (\bibinfo {year} {2004})}\BibitemShut {NoStop}%
\bibitem [{\citenamefont {Mike}\ and\ \citenamefont {Farmer}(2008)}]{Mike2005}%
  \BibitemOpen
  \bibfield  {author} {\bibinfo {author} {\bibfnamefont {Szabolcs}\
  \bibnamefont {Mike}}\ and\ \bibinfo {author} {\bibfnamefont {J~Doyne}\
  \bibnamefont {Farmer}},\ }\bibfield  {title} {\enquote {\bibinfo {title} {{An
  empirical behavioral model of liquidity and volatility}},}\ }\href@noop {}
  {\bibfield  {journal} {\bibinfo  {journal} {Journal of Economic Dynamics and
  Control}\ }\textbf {\bibinfo {volume} {32}},\ \bibinfo {pages} {200--234}
  (\bibinfo {year} {2008})}\BibitemShut {NoStop}%
\bibitem [{\citenamefont {Harris}(2003)}]{harris2003trading}%
  \BibitemOpen
  \bibfield  {author} {\bibinfo {author} {\bibfnamefont {Larry}\ \bibnamefont
  {Harris}},\ }\href@noop {} {\emph {\bibinfo {title} {Trading and exchanges:
  Market microstructure for practitioners}}}\ (\bibinfo  {publisher} {Oxford
  University Press, USA},\ \bibinfo {year} {2003})\BibitemShut {NoStop}%
\bibitem [{\citenamefont {Hendershott}\ \emph {et~al.}(2011)\citenamefont
  {Hendershott}, \citenamefont {Jones},\ and\ \citenamefont
  {Menkveld}}]{hendershott2011does}%
  \BibitemOpen
  \bibfield  {author} {\bibinfo {author} {\bibfnamefont {Terrence}\
  \bibnamefont {Hendershott}}, \bibinfo {author} {\bibfnamefont {Charles~M}\
  \bibnamefont {Jones}}, \ and\ \bibinfo {author} {\bibfnamefont {Albert~J}\
  \bibnamefont {Menkveld}},\ }\bibfield  {title} {\enquote {\bibinfo {title}
  {Does algorithmic trading improve liquidity?}}\ }\href@noop {} {\bibfield
  {journal} {\bibinfo  {journal} {The Journal of Finance}\ }\textbf {\bibinfo
  {volume} {66}},\ \bibinfo {pages} {1--33} (\bibinfo {year}
  {2011})}\BibitemShut {NoStop}%
\bibitem [{\citenamefont {Biais}\ \emph {et~al.}(1995)\citenamefont {Biais},
  \citenamefont {Hillion},\ and\ \citenamefont {Spatt}}]{biais1995empirical}%
  \BibitemOpen
  \bibfield  {author} {\bibinfo {author} {\bibfnamefont {Bruno}\ \bibnamefont
  {Biais}}, \bibinfo {author} {\bibfnamefont {Pierre}\ \bibnamefont {Hillion}},
  \ and\ \bibinfo {author} {\bibfnamefont {Chester}\ \bibnamefont {Spatt}},\
  }\bibfield  {title} {\enquote {\bibinfo {title} {An empirical analysis of the
  limit order book and the order flow in the paris bourse},}\ }\href@noop {}
  {\bibfield  {journal} {\bibinfo  {journal} {the Journal of Finance}\ }\textbf
  {\bibinfo {volume} {50}},\ \bibinfo {pages} {1655--1689} (\bibinfo {year}
  {1995})}\BibitemShut {NoStop}%
\bibitem [{\citenamefont {Schmitt}\ \emph {et~al.}(2012)\citenamefont
  {Schmitt}, \citenamefont {Sch{\"a}fer}, \citenamefont {M{\"u}nnix},\ and\
  \citenamefont {Guhr}}]{Schmitt2012}%
  \BibitemOpen
  \bibfield  {author} {\bibinfo {author} {\bibfnamefont {Thilo~A}\ \bibnamefont
  {Schmitt}}, \bibinfo {author} {\bibfnamefont {Rudi}\ \bibnamefont
  {Sch{\"a}fer}}, \bibinfo {author} {\bibfnamefont {Michael~C}\ \bibnamefont
  {M{\"u}nnix}}, \ and\ \bibinfo {author} {\bibfnamefont {Thomas}\ \bibnamefont
  {Guhr}},\ }\bibfield  {title} {\enquote {\bibinfo {title} {{Microscopic
  understanding of heavy-tailed return distributions in an agent-based
  model}},}\ }\href {?} {\bibfield  {journal} {\bibinfo  {journal} {EPL
  (Europhysics Letters)}\ }\textbf {\bibinfo {volume} {100}},\ \bibinfo {pages}
  {38005} (\bibinfo {year} {2012})}\BibitemShut {NoStop}%
\bibitem [{\citenamefont {M{\"u}nnix}\ \emph {et~al.}(2010)\citenamefont
  {M{\"u}nnix}, \citenamefont {Sch{\"a}fer},\ and\ \citenamefont
  {Guhr}}]{munnix2010impact}%
  \BibitemOpen
  \bibfield  {author} {\bibinfo {author} {\bibfnamefont {Michael~C}\
  \bibnamefont {M{\"u}nnix}}, \bibinfo {author} {\bibfnamefont {Rudi}\
  \bibnamefont {Sch{\"a}fer}}, \ and\ \bibinfo {author} {\bibfnamefont
  {Thomas}\ \bibnamefont {Guhr}},\ }\bibfield  {title} {\enquote {\bibinfo
  {title} {{Impact of the tick-size on financial returns and correlations}},}\
  }\href {?} {\bibfield  {journal} {\bibinfo  {journal} {Physica A}\ }\textbf
  {\bibinfo {volume} {389}},\ \bibinfo {pages} {4828--4843} (\bibinfo {year}
  {2010})}\BibitemShut {NoStop}%
\end{thebibliography}%

\end{document}